\documentclass{aa}  

\usepackage{amsmath}
\usepackage{graphicx}
\usepackage{svg}
\usepackage[normalem]{ulem}
\usepackage{xcolor}
\usepackage{listings}

\definecolor{DarkGreen}{RGB}{0,150,0}
\lstdefinelanguage{ADQL}{
  morekeywords={SELECT, FROM, WHERE, AND, CONTAINS, POINT, CIRCLE, ABS, BETWEEN},
  sensitive=true,
  morecomment=[l]{--},
  morestring=[b]',
}

    \usepackage[colorlinks = true, allcolors=blue]{hyperref}

    \usepackage{orcidlink}
    \usepackage{stfloats}
    \usepackage{pdflscape}

\usepackage{txfonts}
\usepackage{float}
\usepackage{booktabs}
\usepackage{subcaption}

\begin{document}

\title{Tidal tails in open clusters:}
\subtitle{Morphology, binary fraction, dynamics, and rotation}
\titlerunning{Tidal tails in open clusters}

\author{
    Ira Sharma\inst{1,}\inst{2}\orcidlink{0009-0009-0602-0751}
    \and 
    Vikrant Jadhav\inst{3,4}\orcidlink{0000-0002-8672-3300} 
    \and
    Annapurni Subramaniam\inst{2}\orcidlink{0000-0003-4612-620X} 
    \and
    Henriette Wirth\inst{4}\orcidlink{0000-0003-1258-3162}
}

\institute{
    Department of Physical Sciences, IISER Mohali, Knowledge City, Sector 81, SAS Nagar, Manauli PO 140306, Punjab, India\\
    \email{irasharma2208@gmail.com}
    \and
    Indian Institute of Astrophysics, 2nd Block, Koramangala, Bangalore-560034, India\\
    \email{purni@iiap.res.in}
    \and
    Helmholtz-Institut für Strahlen- und Kernphysik, Universität Bonn, Nussallee 14-16, 53115 Bonn, Germany\\
    \email{vjadhav@astro.uni-bonn.de}
    \and
    Charles University, Faculty of Mathematics and Physics, Astronomical Institute, V Hole\v{s}ovi\v{c}kách 2, Praha, CZ-18000, Czech Republic
}

   \date{Received May 28, 2025; accepted October 1, 2025}

   \abstract 
   {This research presents unsupervised machine learning and statistical methods to identify and analyze tidal tails in open star clusters using data from the Gaia DR3 catalog.}
   {We aim to identify member stars and to detect and analyze tidal tails in five open clusters, BH 164, Alessi 2, NGC 2281, NGC 2354, and M67, of ages between 60 Myr and 4 Gyr. These clusters were selected based on the previous evidence of extended tidal structures.}
   {We utilized machine learning algorithms such as Density-based Spatial Clustering of Applications with Noise (DBSCAN) and principal component analysis (PCA), along with statistical methods to analyze the kinematic, photometric, and astrometric properties of stars. Key characteristics of tidal tails, including radial velocity, the color-magnitude diagram, and spatial projections in the tangent plane beyond the cluster's Jacobi radius ($r_J$), were used to detect them. We used N-body simulations to visualize and compare the observables with real data. Further analysis was done on the detected cluster and tail stars to study their internal dynamics and populations, including the binary fraction. We also applied the residual velocity method to detect rotational patterns in the clusters and their tails.}
   {We identified tidal tails in all five clusters, with detected tails extending farther in some clusters and containing significantly more stars than previously reported (tails ranging from 40 to 100 pc, one to four times their $r_J$, with 100-200 tail stars). The luminosity functions of the tails and their parent clusters were generally consistent, and tails lacked massive stars. In general, the binary fraction was found to be higher in the tidal tails. Significant rotation was detected in M67 and NGC 2281 for the first time.}
   {}
   

   \maketitle

\section{Introduction}

The Milky Way, our home galaxy, is a barred spiral galaxy with a central bulge, spiral arms, and a thin and thick disk \citep{2014A&A...569A.125H, Shen_2020}. This complicated structure also contains a dark matter halo, which affects the motions of stars and stellar clusters \citep{2019A&A...621A..56P, 2019MNRAS.490.3426D}. The Milky Way's stellar population is diversified, with old, metal-poor stars in the halo and new, metal-rich stars in the thin disk. This complex stellar environment contains a variety of stellar groups, including open clusters, globular clusters, and tidal streams, with each of which serving as a natural laboratory for studying Galactic dynamics, star formation, and stellar evolution \citep{2012A&A...543A.156K, 2013A&A...558A..53K, 2019MNRAS.484.2832V, 2020ARA&A..58..205H, universe8020111}. 

Tidal tails are extended streams of stars stretching from clusters due to the Milky Way's gravitational effects, bearing the signatures of Galactic dynamics on clusters. For example, tidal tails can reveal a cluster's degree of relaxation and dynamical condition, revealing whether it is still bound or disintegrating. The tidal tails provide important information on the cluster's previous trajectory, dynamics, and mass loss history. The length, shape, and orientation of the tidal tails are determined by the cluster's initial mass, age, and orbit within the Galactic potential. By examining these characteristics, we can piece together the cluster's path across the Milky Way, revealing details on the structure of the Galactic gravitational potential, the distribution of dark matter, Galactic tides \citep{1999A&A...352..149C, 2002MNRAS.332..915I, Montuori_2007}, etc. 
 
The high stellar density, strong gravitational binding, and reduced field star contamination make the tidal tails of globular clusters easier to detect \citep{2000A&A...359..907L, 2001ApJ...548L.165O, Odenkirchen_2003}. The low stellar densities of open clusters and substantial field star contamination from the dense stellar background of the Galactic disk pose significant challenges for detecting the tidal tails of open clusters \citep{2001A&A...377..462B, 2010MNRAS.402.1841C, 2011A&A...531A..92R}. Also, because of internal relaxation, interactions with molecular clouds, and the disruptive effects of the Galactic tidal field, open clusters frequently dissolve over time, leaving their tidal tails poorly populated and difficult to identify.

However, the study of these clusters has been transformed by the advances in astrometric data from the Gaia mission, particularly Data Release 2 \citep{2018A&A...616A...1G} and Gaia EDR3 \citep{2021A&A...649A...1G}. The unprecedented accuracy in proper motion, parallax, and radial velocity measurements allow for the precise separation of field stars from cluster members, greatly increasing the detection of extended features such as tidal tails. These observations have revealed many open clusters with extended star envelopes and tidal tails.

In particular, \citet{2019A&A...622L..13M} discovered the Pisces-Eridanus stream using wavelet decomposition and Density-Based Spatial Clustering of Applications with Noise (DBSCAN), identifying 256 stars in a 400 pc structure. Subsequent work by \citet{2020A&A...639A..64R} and \citet{2020A&A...638A...9R} expanded membership using support vector machines and convergent point methods. \citet{2019AJ....158..122K} and \citet{2020AJ....160..279K} applied hierarchical density-based clustering (HDBSCAN) to Gaia DR2 and identified 1312 stellar groups and 328 stellar strings.
\citet{2021A&A...645A..84M} used Gaia DR2 to identify the tidal tails and coronae in ten open clusters and developed a velocity-based member identification method, starting with the \citet{2018A&A...618A..93C} membership tables.
\citet{2020PASJ...72...47G, 2020ApJ...894...48G} applied principal component analysis (PCA) and Gaussian mixture model (GMM) to identify extra-tidal stars in M67 and NGC 2506, respectively, using Gaia DR2. 

\citet{2022MNRAS.517.3525B} employed ML-MOC, a membership algorithm based on k-nearest neighbors (kNN) and GMM for precise identification from Gaia EDR3 data, and identified 46 open clusters with a stellar corona, while \citet{2022A&A...659A..59T} used  HDBSCAN to find members up to 50 pc from cluster centers, fitted King profiles, used GMMs to analyze spatial distributions, and reported the detection of 71 open clusters with tidal tails.
\citet{2022ApJ...936..160Z} used a combination of Gaia DR2 and Gaia DR3 to investigate the stellar strings identified in \citet{2019AJ....158..122K} by manually linking stellar groups, while \citet{2022AJ....163..275A} analyzed Theia
456 using Gaia EDR3 and produced a more robust membership catalog of 362 stars using DBSCAN.
\citet{2021A&A...647A.137J} used the compact convergent point method to recover 800 pc long tidal tails of the Hyades star cluster using Gaia DR2 and EDR3.

Using Gaia DR3, \citet{Hunt2023A&A...673A.114H} applied the HDBSCAN algorithm to recover 7167 clusters, also including tidal tails for some clusters. However, \citet{2024A&A...686A..42H} derived completeness-corrected photometric masses for 6956 clusters from the earlier work and used these masses to compute their Jacobi radius ($r_J$). \citet{2024A&A...691A..28K} simulated cluster dissolution to estimate the likelihood of star positions, comparing these with Gaia DR3 data to refine membership probabilities. Recently, \citet{Risbud2025arXiv250117225R} used the convergent point method to identify tidal tails in 19 open clusters within 400 pc.

The previous methods often struggled with limitations such as poor astrometric precision, model dependencies, or applicability limited to nearby clusters. These constraints significantly reduce the number of clusters that can be studied in detail. To address this, we designed a method that uses optimized clustering and kinematic analysis to identify tidal tails even in distant systems. In this work, we have used the Gaia DR3 dataset to identify and investigate tidal tails in five open clusters: BH 164, NGC 2354, Alessi 2, NGC 2281, and M67. We tuned the algorithms to increase the detected extent of tidal tails, incorporating radial velocity (RV) filtering (where available), color-magnitude diagram (CMD) matching for age and composition consistency, and an unsupervised approach using PCA to determine the principal axis of elongation, followed by DBSCAN for tail detection. The identified tails were validated using the Jacobi limit to confirm their tidal origin. Projection effects were taken into account in the tangential plane, providing a clearer spatial distribution of the tails. We detected a greater number of stars in the four clusters and identified longer tidal tails in three of them compared to previous studies. Additionally, we obtained a more accurate estimate of the projected tail extent for M67, whose orientation aligns well with the expected orbit, consistent with N-body simulations.
Furthermore, rotation was studied in the M67 and NGC 2281 clusters and tails. The residual velocity method, adopted from \citet{2022ApJ...938..100H, 2024ApJ...963..153H}, was previously used exclusively for member stars to confirm internal rotation in clusters such as Praesepe, Pleiades, Alpha Persei, and Hyades. In this study, we applied it to the cluster and its tails to analyze their rotational patterns.

Section \ref{sec:2} outlines the data preparation, identification of the core cluster, and also explains the N-body simulations used for reference. Section \ref{sec:3} presents the methods used to detect tidal tails. In Sect. \ref{sec:4}, we present a detailed analysis utilizing proper motion vector analysis, orbit integration, luminosity functions, and the binary fractions to explore the cluster and tail dynamics. Section \ref{sec:5} presents our tidal tail findings and discusses the results. In Sect. \ref{sec:6}, the rotational aspect of M67 and NGC 2281 is explored using the residual motion of stars to study the presence of three-dimensional rotation. Finally, we discuss and summarize our research in Sects. \ref{sec:7} and \ref{sec:8}, respectively.

\section{Data and cluster identification}
\label{sec:2}
This section outlines the methods for identifying clusters using Gaia data, including data preparation and clustering techniques to accurately separate cluster members from field stars.

\subsection{Cluster selection}
To maximize computing efficiency, we restricted ourselves to open clusters with manageable dataset sizes spanning up to 10° in radius (i.e., 20° diameter) around each cluster center. We chose five open clusters: BH 164, Alessi 2, NGC 2281, NGC 2354, and M67, as shown in Table \ref{tab:cluster_parameters}. \citet{2022MNRAS.517.3525B} detected tidal structures in BH 164 and Alessi 2, while \citet{2022A&A...659A..59T} and \citet{Hunt2023A&A...673A.114H, 2024A&A...686A..42H} found evidence of tidal tails in NGC 2281 and NGC 2354. Given their reported tidal structures, these clusters provided good test cases for validating our detection method and confirming the presence of their tidal tails. The inclusion of M67, which previously lacked recognized tidal tails, was justified by the observations of \citet{2020PASJ...72...47G}, indicating a potential for such features.

The age spectrum of the selected clusters ranges from 60 million to 1.5 billion years, with M67 of around 4 billion years of age. 
The distances vary between 400 and 1300 parsecs, allowing us to investigate how tidal features appear in various Galactic contexts. 

\subsection{Gaia DR3}

The latest Gaia Data Release 3 (DR3; \citealt{2023A&A...674A...1G}) provides a substantial update over Gaia DR2, contains a total of 34 months of observational data, and features major advancements in data processing techniques. 
Gaia DR3 catalogs 1.47 billion sources with 5- or 6-parameter astrometric solutions, achieving a 30\% improvement in parallax precision and nearly doubling the accuracy of proper motion measurements. These enhancements significantly improve the visibility of open clusters within the Gaia dataset, particularly in proper motion diagrams. The signal-to-noise ratio (S/N) for identifying distant open clusters has increased by approximately four \citep{Hunt2023A&A...673A.114H}, benefiting from a more compact Gaussian distribution of cluster stars for those with proper motion dispersions now within the reduced Gaia error margins.  

For reference, the frequently used Gaia DR3 quantities in this work are:  
\begin{itemize}
    \item ra, dec ($\alpha$ and $\delta$ in deg): celestial coordinates in the ICRS frame. Typical errors: $\approx$0.01–0.1 mas.  
    \item pmra, pmdec ($\mu_\alpha \cos\delta$ and $\mu_\delta $ in mas yr$^{-1}$): proper motions in ra and dec, respectively. Typical errors: $\approx$0.02–0.2 mas yr$^{-1}$. 
    \item parallax ($\varpi$ in mas): used for distance estimation. Typical errors: $\approx$0.02 mas for bright stars, $\sim$0.5 mas at $G \approx 20$.  
    \item distance (in pc): calculated as 1000 / $\varpi$.  
    \item G magnitude (in mag): Gaia broad-band photometry. Errors: $\lesssim$0.001 mag for bright stars, $\approx$0.02 mag at $G \approx 20$.  
    \item BP–RP (in mag): color index from Gaia blue and red photometers. Typical errors: $\lesssim$0.01 mag (bright), $\approx$0.05 mag (faint).  
    \item $M_G$ (in mag): absolute $G$ magnitude, derived from $G$ and $\varpi$.  
    \item RV (in km s$^{-1}$): line-of-sight velocity, available for bright stars ($G \lesssim 15$). Typical errors: $\approx$0.2–5 km s$^{-1}$.  
\end{itemize}
\begin{figure*}[t]
\centering
\includegraphics[width=0.93\textwidth, height=0.21\textheight]{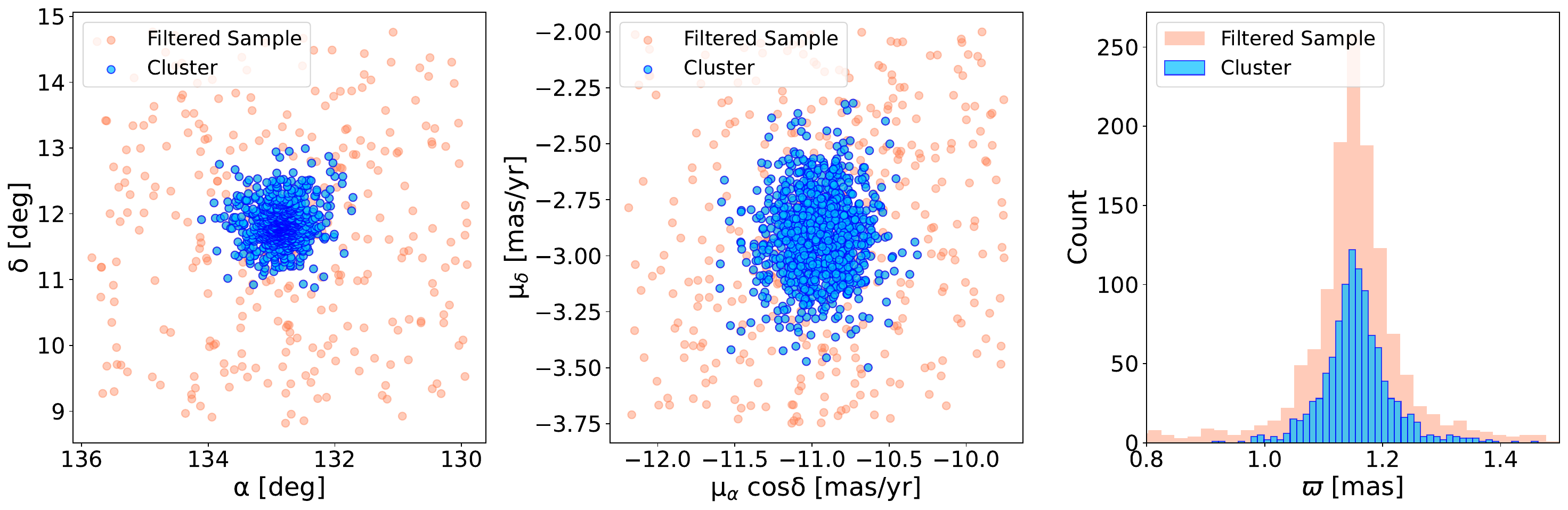} 
\caption{Spatial distribution, proper motion, and parallax of the filtered sample (light coral) after filtering and the cluster members (blue) after applying DBSCAN for M67.}
\label{fig:1}
\end{figure*}

\subsection{Data filtering and preparation}
The dataset was refined using the following steps to ensure high-quality, accurate data suited to star cluster analysis.

\begin{itemize}
 
\item Proper motion filtering:  
pmra\_min $\leq$ pmra $\leq$ pmra\_max and \quad pmdec\_min $\leq$ pmdec $\leq$ pmdec\_max.

For example, for M67, the ranges used were:  
-12.5 \text{mas yr$^{-1}$} $\leq$ pmra $\leq$ -9.5 \, \text{mas yr$^{-1}$}, \quad -4  \text{mas yr$^{-1}$} $\leq$ pmdec $\leq$ -1  \text{mas yr$^{-1}$}. These bounds were chosen based on the literature values \citep{2020PASJ...72...47G} and by visually inspecting the star distribution in proper motion space.

\item Error filtering: Stars with proper motion errors \(< 0.5\) mas yr$^{-1}$ were retained. We limited the parallax using $\varpi>0$ and $\sigma_{\varpi}/\varpi<0.1$. 
Stars passing these filters were labeled as the filtered sample. We used relative parallax error to maintain consistent distance precision across clusters, although this imposes stricter absolute error thresholds on distant clusters, potentially biasing against the inclusion of faint members. Additionally, unresolved binaries may introduce astrometric errors not explicitly accounted for in this analysis. The spatial, proper motion, and parallax distributions for M67 are highlighted in light coral in Fig. \ref{fig:1}.

\item Data preparation: For core cluster identification, we used the astrometric features: ra, dec, pmra, pmdec, and distance. We used distance instead of parallax to linearize spatial separations, enabling meaningful Euclidean distances in the clustering feature space. Since we filtered stars to have a relative parallax error $< 0.1$, these distances are considered reliable. Stars with missing (NaN) values in any of these features were excluded. These core features were then normalized to zero mean and unit variance to ensure each contributed equally to the clustering process.
\end{itemize}

\subsection{DBSCAN clustering for core cluster identification}
Density-Based Spatial Clustering of Applications with Noise \citep{10.5555/3001460.3001507, 2011JMLR...12.2825P} is a density-based clustering algorithm that identifies groups of closely packed points while treating outliers as noise. Unlike methods such as $k$-means, DBSCAN does not require specifying the number of clusters in advance and can discover clusters of arbitrary shape, making it well-suited for astrophysical data where the boundaries of a star cluster may not be sharply defined. It gives clear-cut clusters without soft membership boundaries.

For a data point to be considered as a part of a cluster, it must have a minimum number of neighboring data points within a specified distance, $\epsilon$. Key Parameters of DBSCAN:
\begin{itemize}
    \item \texttt{$\epsilon$}: Maximum distance for two points to be considered neighbors. 
    \item \texttt {min\_neighbors}: Minimum number of neighboring data points required to form a cluster.
\end{itemize}
The DBSCAN algorithm identifies core points, border points, and noise points. Core points have sufficient neighbors within $\epsilon$; border points are neighbors of core points but are not core themselves; noise points are neither.
In this work, DBSCAN was applied to the scaled 5D feature (ra, dec, pmra, pmdec, distance) space of the filtered sample to identify a dense overdensity corresponding to the core cluster population. For each cluster, the parameters ($\epsilon$, \texttt{min\_neighbors}) were tuned to find the combination that resulted in the most spatially complete and centrally concentrated core in ra–dec space, while minimizing contamination. In M67, this method isolated 1037 cluster members (N$_{memb}$), whose spatial, proper motion, and parallax distributions are also shown in blue in Fig. \ref{fig:1}.

\subsection{N-body simulations for reference}

We used N-body simulations to visualize and compare the observables with real data. We used \textsc{McLuster} \citep{2011MNRAS.417.2300K} to initialize a clusters with 200--3200 M$_{\odot}$ with Plummer profile, no initial mass segregation, half mass radii according to the $M_{ecl}$-$R_h$ relation ($R_h=0.1 (M_{ecl}/M_{\odot})^{0.13} \text{ pc}$; \citealt{2012A&A...543A...8M}), inclusion of Milky Way potential (\texttt{MWPotential2014}; \citealt{2015ApJS..216...29B}), position/velocity similar to the Sun, particle masses according to Kroupa initial mass function \citep{2001MNRAS.322..231K}, no primordial binaries, and solar metallicity. The dynamical evolution was performed using \textsc{PeTar} \citep{2020MNRAS.497..536W} and the stellar evolution using the \textsc{bse} code \citep{2013ascl.soft03014H}. See \cite{Jadhav2025_tidal_tail_ii} for more details.

We used a snapshot with the age and mass comparable to the cluster sample. Rotation around the Galactic center was performed to orient the tidal tails along the cluster orbit correctly. As the simulated clusters have circular orbits and slightly different Galactocentric radii than the clusters, minor translations were done to match the 6D position of the simulated cluster with the real cluster. 
Synthetic observations of these clusters were done using \textsc{astropy} \citep{2013A&A...558A..33A, 2018AJ....156..123A, 2022ApJ...935..167A} to visualize the theoretically expected observables of real clusters (e.g., sky distribution and proper motions).

\section{Tidal tail detection}
\label{sec:3}
Once core members were identified, we searched for tidal tail stars among noncluster filtered sample stars within 500 pc of the cluster’s mean distance. The approach combined RV filtering, CMD matching, and machine learning with \texttt{scikit-learn} \citep{2011JMLR...12.2825P}, including k-d tree (k-dimensional tree), PCA, and DBSCAN. Tidal stars are expected to share RVs and CMD features with the cluster, align along the cluster's elongation axis, and lie beyond their Jacobi limit.

\begin{table*}
\centering
\small
\caption{\label{tab:cluster_parameters}Core cluster parameters of the selected open clusters.}
\begin{tabular}{lccccccccccc}
    \hline\hline
    Name & ra & dec & $l$ & $b$ & Age & pmra & pmdec & $\varpi$ & Dist & RV & $N_{\rm memb}$\\
     & (deg) & (deg) & (deg) & (deg) & log$_{10}$(yr) & (mas yr$^{-1}$) & (mas yr$^{-1}$) & (mas) & (pc) & (km s$^{-1}$) & \\
    \hline
    BH 164 & 222.20 & $-66.45$ & 314.30 & $-6.19$ & 7.81 & $-7.41$ & $-10.69$ & $2.40 \pm 0.10$ & 419 & $-3.4 \pm 3.3$ & 289 \\
    Alessi 2 & 71.61 & 55.16 & 152.38 & 6.36 & 8.76 & $-0.92$ & $-1.08$ & $1.63 \pm 0.05$ & 616 & $-10.3 \pm 4.2$ & 130\\
    NGC 2281 & 102.09 & 41.05 & 174.94 & 16.88 & 8.78 & $-2.96$ & $-8.27$ & $1.94 \pm 0.07$ & 517 & $19.6 \pm 4.6$ & 438 \\
    NGC 2354 & 108.49 & $-25.73$ & 238.38 & $-6.87$ & 9.15 & $-2.86$ & $1.86$ & $0.78 \pm 0.03$ & 1284 & $31.8 \pm 5.5$ & 244\\
    M67 & 132.85 & 11.82 & 215.68 & 31.92 & 9.56 & $-10.97$ & $-2.92$ & $1.16 \pm 0.04$ & 867 & $33.9 \pm 3.4$ & 1037 \\
    \hline
\end{tabular}
\tablefoot{
Col. 1: Cluster name; Cols. 2–3: Equatorial coordinates; Cols. 4–5: Galactic lon-lat; Col. 6: Age from \citet{2019A&A...623A.108B} and \citet{2020A&A...640A...1C}; Cols. 7-10: Proper motions, mean parallax, and mean distance of identified members; Col. 11: Mean radial velocity of identified core cluster members; Col. 12: Number of identified core cluster members.
}
\end{table*}

\subsection{Radial velocity filtering}
\label{subsec:rvfiltering}
After defining the core cluster, we applied an RV filter to further refine the potential tail candidates. Specifically, we selected stars within the 500 pc distance range whose RVs fall within a range of $\pm 10$ km s$^{-1}$ of the core cluster’s mean RV:
\begin{equation}
\texttt{RV}_{\text{tail}} = \texttt{RV}_{\text{core}} \pm 10 \, \text{km s$^{-1}$}.
\end{equation}
This $\pm 10$ km s$^{-1}$ threshold serves as a loose filtering criterion to remove clear RV outliers that are inconsistent with the cluster’s dynamics. It is broad enough to account for small projection-induced velocity gradients\footnote{Projection effects are corrected after the tidal tail detection for further analysis; see Sect. \ref{subsec:correction}.}, which arise from geometric effects over large angular separations. This window was chosen to preserve plausible tail candidates while excluding clear outliers.
Since Gaia DR3 provides RVs only for a subset of stars, this filtering step is applied only where RV data is available. For stars lacking RV data, we rely on CMD-based filtering to evaluate their consistency with the cluster’s stellar population.

\subsection{Color-magnitude diagram matching}
In addition to RV, we used the color (BP-RP) versus apparent G magnitude CMD to refine our identification of tidal tail stars. In the CMD space, tidal tail stars are expected to closely resemble the core cluster members, sharing similar age, metallicity, and evolutionary history. The CMD provides a visual representation of these properties and allows us to distinguish between stars that formed together and those that are unrelated to background or foreground stars.

To efficiently process large datasets such as Gaia DR3, we employed a k-d tree structure for performing the CMD cut. This algorithm enables fast nearest-neighbor searches in high-dimensional space and was used to identify stars whose CMD values closely match those of the core cluster members. The tree partitions the data recursively
by splitting on dimensions such as color and magnitude. A distance threshold of 0.05 in the CMD space was applied to select potential tidal tail members, ensuring photometric similarity with the cluster core. The CMD distance is defined as
\begin{equation}
d_{\text{CMD}} = \sqrt{(G - G_{\text{core}})^2 + \left[(BP - RP) - (BP_{\text{core}} - RP_{\text{core}})\right]^2}.
\end{equation}
Stars lying within \( d_{\text{CMD}} \leq 0.05 \) mag of each cluster CMD point (including unresolved binaries and post-main sequence stars) were considered as tail candidates.

\subsection{Principal component analysis and DBSCAN clustering}
Principal component analysis is a dimensionality reduction technique that identifies the principal axes of variance in a dataset. In the context of tidal tail detection, PCA helped to determine the primary direction of elongation for the core cluster, aiding in the analysis of the spatial distribution of stars.
\begin{itemize}
 \item Principal axis determination: The positions of the core cluster stars were used to perform PCA, identifying the principal axis along which the cluster is elongated. This axis is crucial for transforming the positions of surrounding stars to align with the cluster’s spatial orientation.
\item Transformation: Stars in the surrounding field were transformed based on the principal axis to create a new coordinate system that emphasized the elongated structure of the tidal tails.
\end{itemize}
The DBSCAN algorithm was then applied to the transformed spatial data to identify clusters corresponding to the tidal tail. In this application, DBSCAN differentiated between stars in the tidal tails and noise (field stars).

\begin{figure*}[t]
\centering
\includegraphics[width=0.87\textwidth, height=0.46\textheight]{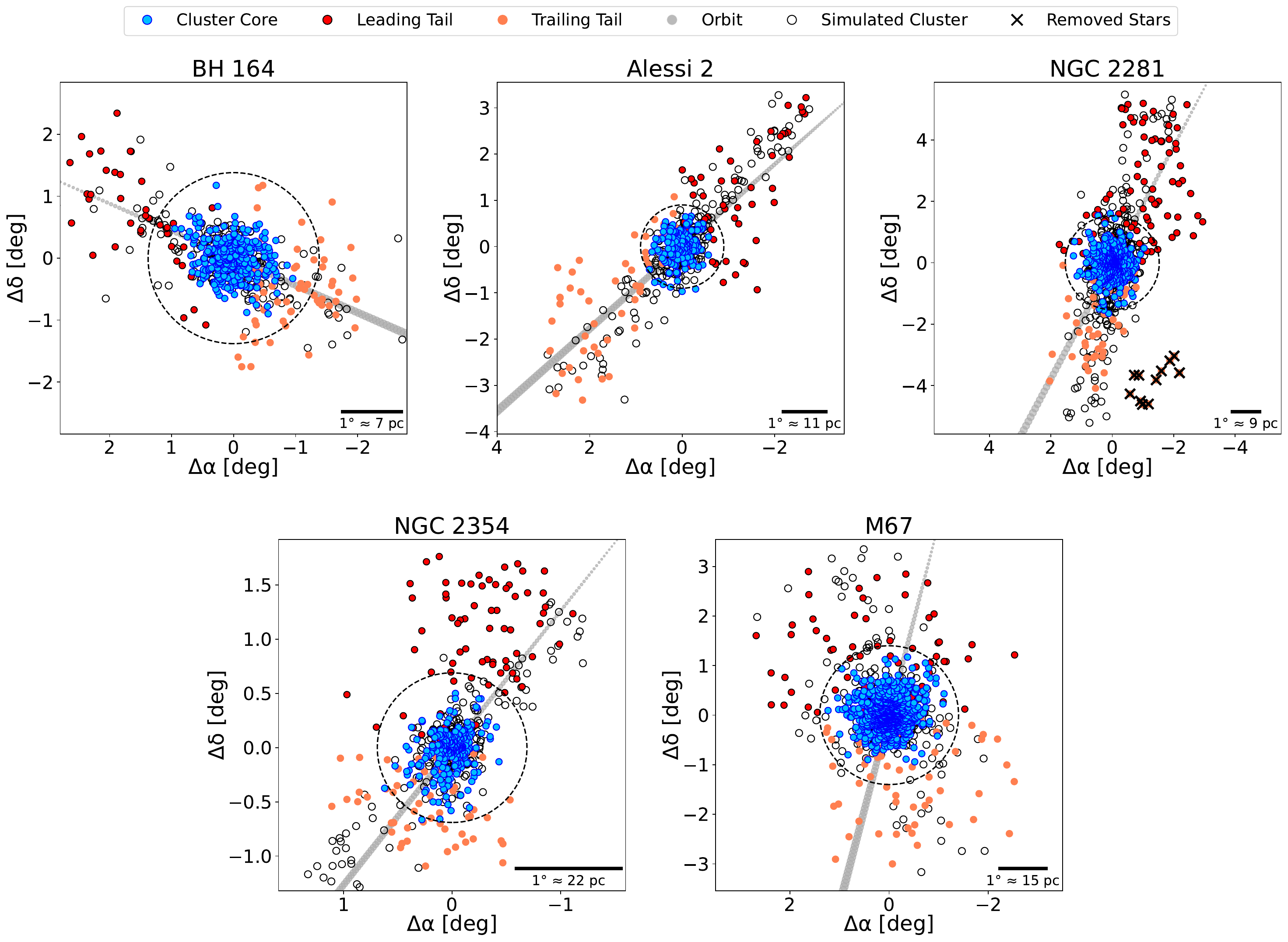} 
\caption{Spatial distribution of the identified core members (blue dots) for the five clusters with their detected tidal tails (red and coral dots), N-body simulated cluster (empty dots), Jacobi limit (dotted black circle), and expected orbit (gray arrow) pointing toward the direction of motion (refer to Subsect. \ref{subsec:orbit}). Crosses indicate manually removed stars for NGC 2281. }
\label{fig:2}
\end{figure*}
\subsection{Tangential projection of the celestial coordinates}
\label{subsec:tangential}
We employed a tangential projection of celestial coordinates onto a 2D Cartesian plane to analyze the spatial distribution of tidal tails in star clusters. This approach simplified visualization and enabled accurate spatial analyzes by mitigating the effects of spherical geometry. 

For the trigonometric transformation, right ascension ($\alpha$) and declination ($\delta$) are first converted to radians, as is the reference point $(\alpha_0, \delta_0)$, corresponding to the mean position of the cluster core. The offsets in the tangent plane are computed as follows:
\begin{align}
    \Delta \alpha &= \cos(\delta) \sin(\alpha - \alpha_0), \\
    \Delta \delta &= \sin(\delta) \cos(\delta_0) - \cos(\delta) \sin(\delta_0) \cos(\alpha - \alpha_0).
\end{align}
The trigonometric expressions are dimensionless, but when interpreted under the small-angle approximation, they correspond to angular offsets expressed in radians. For visualization in Fig. \ref{fig:2}, these offsets were explicitly converted to degrees.
Although the projection provides a convenient 2D representation, it inevitably reduces information from the full 3D structure.

\subsection{Beyond the Jacobi radius}
The final criterion for checking our detection of the tidal tail was the spatial location of the stars relative to the best-fit $r_J$ values for each cluster as provided in Table \ref{tab:tail_parameters} taken from \citet{2024A&A...686A..42H}, in which $r_J$  is defined as the distance from the cluster’s center to its $L_1$ Lagrange point, effectively marking the boundary beyond which stars are no longer bound to the cluster. Their full formulation accounts for the cluster mass as well as its orbital parameters around the host galaxy, including the circular frequency ($\Omega$) and the epicyclic frequency ($k$).

We note that $r_J$ is not used as a strict boundary in Fig. \ref{fig:2} for defining tidal tail stars. Due to projection effects in the sky plane, some tail stars may still appear within the projected $r_J$. Our use of this limit is only to show that the detected features extend beyond the theoretical boundary.
\begin{table*}[t]
\centering
\small
\caption{\label{tab:tail_parameters}Parameters of the five open clusters and their tidal tails.}
\begin{tabular}{lcccccccccccc}
\hline\hline
Name & X & Y & Z & $r_J$ & $L_{\mathrm{lead}}$ & $L_{\mathrm{trail}}$ & $N_{\mathrm{lead}}$ & $N_{\mathrm{trail}}$ & $BF_{\mathrm{cluster}}$ & $BF_{\mathrm{lead}}$ & $BF_{\mathrm{trail}}$ \\
 & (pc) & (pc) & (pc) & (pc) & (pc) & (pc) & & & & & \\
\midrule
BH 164 & -7830.95 & -298.14 & -25.32 & 9.90 & 12.98 & 6.74 & 46 & 61 & $0.26\pm0.04$ & $0.19\pm0.07$ & $0.33\pm0.09$ \\
Alessi 2 & -8664.69 & 284.62 & 90.45 & 9.55 & 35.97 & 35.13 & 51 & 49 & $0.22\pm0.05$ & $0.54\pm0.13$ & $0.40\pm0.11$ \\
NGC 2281 & -8615.26 & 43.48 & 172.18 & 13.58 & 38.50 & 25.34 & 92 & 44 & $0.24\pm0.03$ & $0.24\pm0.06$ & $0.31\pm0.10$ \\
NGC 2354 & -8794.02 & -1088.29 & -131.38 & 14.90 & 25.05 & 14.18 & 82 & 60 & $0.34\pm0.06$ & $0.23\pm0.07$ & $0.31\pm0.09$ \\
M67 & -8718.53 & -428.90 & 481.10 & 20.56 & 29.45 & 31.36 & 74 & 89 & $0.28\pm0.02$ & $0.33\pm0.09$ & $0.26\pm0.08$ \\
\bottomrule
\end{tabular}
\tablefoot{Col. 1: Cluster name; Cols. 2–4: Galactocentric positions of the identified members;  Col. 5: Best-fitting $r_J$ from \citet{2024A&A...686A..42H}; Cols. 6–7: Approximate projected extent of the tidal tails beyond $r_J$ determined using the angular separation from the cluster center and the cluster distance, with the orbit direction aiding in identifying the tail components (Subsect. \ref{subsec:orbit}); Cols. 8-9: Number of stars in leading and trailing tails; Cols. 10-12: Binary fractions (for mass ratio $>0.5$) in the cluster, leading tail and trailing tail (Subsect. \ref{subsec:4.4}).}
\end{table*}

\section{Tidal tail analysis}
\label{sec:4}
We analyzed the kinematic properties of each cluster and its tidal tails using proper motion vectors and orbit integration. Proper motions were used to characterize relative stellar motions and identify subtle patterns, while orbit integration traced past and future trajectories within the Milky Way’s potential. We also determined and analyzed the luminosity functions and the binary fractions for both clusters and their tails.
\subsection{Correction of proper motion vectors for projection effects}
\label{subsec:correction}

Projection effects influence the apparent proper motions and RVs of individual stars within a star cluster due to the relative position and motion of the cluster with respect to the Sun \citep{2009A&A...497..209V, 2024A&A...687A..89J}. These effects arise because of the angular separation of stars from the cluster center and the cluster's bulk motion. Neglecting these corrections can lead to artificial signatures, such as false rotation or contraction within the cluster. Correcting these projection-induced deviations is essential for accurately analyzing the intrinsic motion of stars.

To account for this, we corrected both the proper motions and RVs of stars by removing the projection-induced components, which depend on the cluster’s bulk motion, the star's angular offset from the cluster center, and the parallax values of both the star and the cluster. The mathematical formalism for this correction is detailed in \citet{2009A&A...497..209V}. The resulting corrected proper motions better reflect the intrinsic kinematics of the cluster.
\subsection{Orbit integration}
\label{subsec:orbit}
Using the observed position and velocity data of the cluster, we computed its Galactic orbit employing the \texttt{Astropy} and \texttt{Galpy} libraries \citep{2013A&A...558A..33A, 2015ApJS..216...29B, 2018AJ....156..123A, 2022ApJ...935..167A}. The orbit integration was carried out using the widely adopted \texttt{MWPotential2014} model, which represents the Milky Way's gravitational potential through a combination of an exponential disk, a spherical bulge, and a dark matter halo. We adopted the solar position \((-7999.986, 0, 15~\mathrm{pc})\) and velocity parameters \((10, 235, 7~\mathrm{km\,s^{-1}})\) provided in \citet{2015ApJS..216...29B}.

Initial conditions for the orbit were defined using the cluster's ra, dec, distance, proper motions, and RV. The integration process involved numerically solving the equations of motion within the adopted Galactic potential. The resulting orbit was projected onto the tangent plane (the gray arrow in Fig. \ref{fig:2} and \ref{fig:3}) to visualize the cluster's motion and identify the direction of its tidal tails. The leading tail points in the direction of orbital motion, while the trailing tail extends opposite to it.
\subsection{Luminosity function}
\label{subsec:luminosity}
The luminosity functions (LFs) for each cluster and its tidal tails were constructed to facilitate comparison, which can be seen in the subplot (h) of plots \ref{fig:A1} to \ref{fig:A5} in the appendix. Specifically, LFs were generated for two distinct regions: the identified core cluster and the detected tidal tails, including the leading tail and the trailing tail, each using a consistent bin size $M_G$ of 1 mag. The faintest one or two bins in both regions for every cluster contained fewer counts, probably due to observational incompleteness and stellar evaporation. These bins were excluded from the linear fit, which was performed using the following relation:
\begin{equation}
    \log_{10} N \propto a \times M_G,
\end{equation}
where $N$ represents the number of stars in each absolute magnitude $M_G$ bin and $a$ denotes the slope of the LF.

\subsection{Binary fraction}
\label{subsec:4.4}
The binary fractions in the cluster and tails were calculated by identifying unresolved binaries in the Gaia CMD similar to \citet{Jadhav2021AJ....162..264J}. The best-fitting parsec isochrones \citep{Bressan2012MNRAS.427..127B} were obtained based on literature parameters \citep{2019A&A...623A.108B, 2020A&A...640A...1C}. An empirical main sequence ridge line was estimated from the Gaia CMD and the isochrone using \textsc{robustgp} \citep{Li2020ApJ...901...49L, Li2021A&C....3600483L}. Mass ratios of stars were calculated based on the shift away from the main sequence. The sources with mass ratios of more than 0.5 were classified as binary.
The ratio of binaries to the total sources is noted as the binary fraction. A detailed description of the binary fraction calculations will be presented in \citet{Jadhav2025_tidal_tail_ii}.
\begin{figure*}[t]
\centering
\includegraphics[width=0.87\textwidth, height=0.46\textheight]{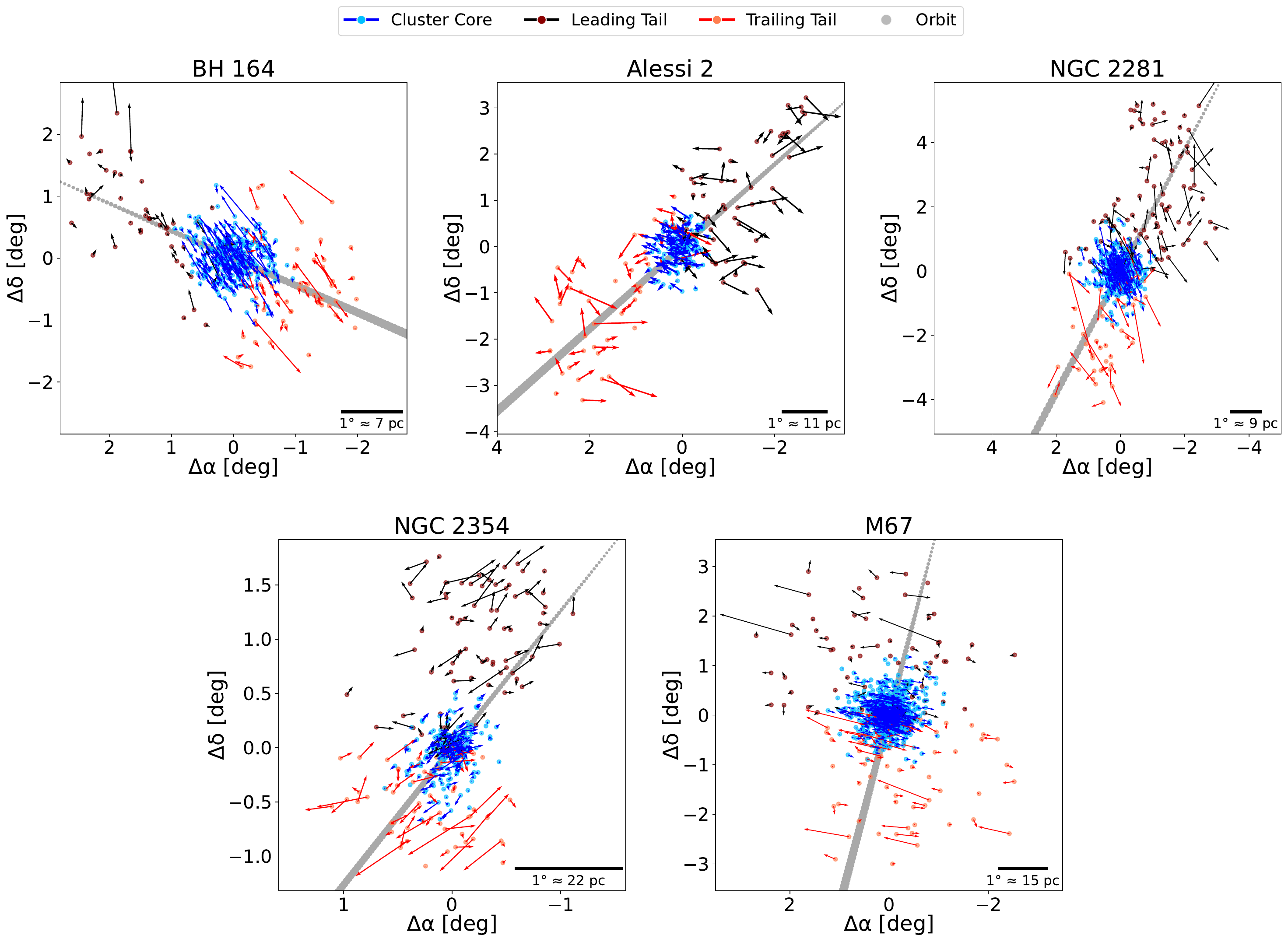} 
\caption{First-order corrected proper motion vectors for the five open clusters.}
\label{fig:3}
\end{figure*}
\section{Tidal tails and their properties}
\label{sec:5}
This section presents results for the five clusters, focusing on tidal tails, internal dynamics, and orbits. We examined mass segregation, spatial and kinematic distributions, binary populations, and tail morphology. To compare the tail and core populations, we applied the Kolmogorov–Smirnov (KS) test, reporting both the KS statistic and p-value: low p-values ($<0.05$) indicated significant differences, while high values suggested the samples are drawn from the same population. We also compared the observed properties with theoretical expectations and prior studies.

\subsection{BH 164 (Alessi 7, MWSC 2255, vdBergh-Hagen 164)}
\citet{2022MNRAS.517.3525B} identified 286 stars in BH 164, while our analysis found 396 stars within the cluster and its two tidal tails, which exhibit an S-shaped structure. The shape and orientation of the leading and trailing tails align with the theoretical N-body simulations (Fig. \ref{fig:2}), and the extent is consistent with \citet{2022MNRAS.517.3525B}. Our study incorporated projection effects, identifying 289 cluster members and 107 tail stars, with the projected span of approximately 40 pc. In Fig. \ref{fig:3}, the proper motion vectors for cluster stars (blue) and tail stars (black and red) reveal a bifurcation pattern around the expected orbit, with opposing motions in the tails indicating outbound velocities. 

Figure \ref{fig:A1} (h) compares the LFs of the cluster (blue) and its tidal tails (red). The KS statistic of 0.10 and a p-value of 0.50 indicate no significant difference between the two distributions, implying that the tails contain a stellar population similar to the cluster. However, the cluster’s LF shows a steeper rise at the bright end, suggesting it retains a larger number of massive stars, while the tails are more populated by lower-mass stars. The binary fraction of the cluster was found to be $0.26 \pm 0.04$, and the same average fraction of 0.26 is found for the tails. However, the binary fractions in the leading ($0.19 \pm 0.07$) and trailing ($0.33 \pm 0.09$) tails suggest an asymmetric ejection of binaries.

\subsection{Alessi 2 (LeDrew 4, MWSC 419)}
\citet{2022MNRAS.517.3525B} identified 201 stars in Alessi 2, whereas our analysis found 230 stars spanning the cluster and its two tidal tails, forming a subtle \reflectbox{S}-shaped structure along the expected orbit. The morphology and orientation are consistent with the theoretical predictions, but the N-body simulations suggested subtle S-shaped tails (Fig. \ref{fig:2}). After accounting for projection effects, we identified 130 core members and 100 tail stars, with a tip-to-tip extent of approximately 90~pc, significantly longer than the 47~pc reported by \citet{2022MNRAS.517.3525B}. Figure \ref{fig:3} shows that the tail stars exhibit no directional motion, with vectors pointing in various directions.
The velocity vectors within the core appear to indicate motion predominantly in one direction, almost perpendicular to the expected orbit. However, upon closer examination, many of these vectors actually represent small motions in various directions, which are not readily visible in the projected plane plot of Fig. \ref{fig:3}.

Figure \ref{fig:A2} (h) compares the LFs of the cluster and its tidal tails, with a KS statistic of 0.29 and p-value of 0.12. While the p-value suggests the distributions are not significantly different, the relatively high statistic points to a noticeable shift: the cluster retains higher-mass stars, while lower-mass stars dominate the tails. The binary fraction in the cluster was found to be $0.22 \pm 0.05$, while the tidal tails exhibited a higher average fraction of $0.47$, with the leading and trailing tails having the binary fractions of $0.54 \pm 0.13$ and $0.40 \pm 0.11$, respectively.

\subsection{NGC 2281 (Collinder 116, MWSC 989, Melotte 51, OCL 446, Theia 635)}
\citet{2022A&A...659A..59T} identified 443 stars in NGC 2281 with a tidal tail semi-major axis of $\approx$21~pc. \citet{Hunt2023A&A...673A.114H, 2024A&A...686A..42H} reported extended features as a byproduct of their analysis with projected extents based on our criteria of $\approx$34~pc and $\approx$19~pc for the leading and trailing tails, respectively, with a total tip-to-tip span of $\approx$80~pc.
The observed tail orientation is similar to the previous findings of \citet{Hunt2023A&A...673A.114H, 2024A&A...686A..42H}, but we additionally identified a fragmented clump in the trailing tail. This clump at (-2, -4) deg in NGC 2281 appeared to be separated from the rest of the tail candidates in both 2D and 3D space. The location is also different from what is expected from the simulations, and these stars are likely contaminants due to their comoving nature. Hence, we have manually removed these stars (crossed out in Fig. \ref{fig:2}). Our analysis after removing the clump detected 574 stars, including 438 cluster members and 136 tail stars. The leading and trailing tails span $\approx$39~pc and $\approx$25~pc, respectively, forming an \reflectbox{S}-shaped structure along the orbit, whose morphology and orientation aligned with the simulations (Fig. \ref{fig:2}). The total tip-to-tip extent came to $\approx$91~pc. Proper analysis of motion vectors (as shown in Fig. \ref{fig:3}) indicates that the leading and trailing tail vectors show similar motion with arrows pointing in opposite directions, leading to the elongation of tails. The motion of the cluster and the tail vectors are slightly angled relative to the direction of the orbital path.

Figure \ref{fig:A3} (h) presents a KS statistic of 0.11 and a high p-value of 0.46, indicating that the cluster and tail populations are statistically similar. Nonetheless, the tails exhibit a slight excess of low-mass stars, while the core retains more high-mass members, consistent with the expected effects of dynamical stripping and mild mass segregation.
The binary fraction was found to be $0.24 \pm 0.03$ in the cluster and an average of $0.28$ in the tails, with the leading and trailing tails having $0.24 \pm 0.06$ and $0.31 \pm 0.10$, respectively.

\subsection{NGC 2354 (Collinder 131, ESO 492-6, FSR 1291, MWSC 1148, OCL 639, Theia 5939)}

\citet{2022A&A...659A..59T} detected 337 stars in NGC 2354 with tidal tails having a semi-major axis of $\approx$24~pc. The extended features reported by \citet{Hunt2023A&A...673A.114H, 2024A&A...686A..42H} based on our projected extent criteria span $\approx$22~pc (leading) and $\approx$9~pc (trailing), with a total tip-to-tip span of $\approx$61~pc.
Our analysis identified 386 stars, including 244 cluster members and 142 tail stars. The leading and trailing tails span $\approx$25~pc and $\approx$14~pc, respectively, with an overall tip-to-tip extent of $\approx$69~pc. The observed leading tail showed a local deviation, making it look \reflectbox{S} shaped. The orientation and morphology of the tidal tails in NGC 2354 (Fig. \ref{fig:2}) are consistent with the features reported by \citet{Hunt2023A&A...673A.114H, 2024A&A...686A..42H}. The leading tail near (0, 1.5) deg extended farther than predicted by the simulations, but in the absence of a clear discontinuity or any distinct break in the stellar distribution, we could not rule out these stars as we did for NGC 2281. There is a possibility these tail candidates are false positives, and caution is advised before using the tails in NGC 2354. Proper motion vectors (Fig. \ref{fig:3}) reveal noncoherent and minimal motion in the cluster core as compared to the tails, which show large motion. The leading tail shows motion in the expected orbit's direction, while the trailing tail moves in the reverse direction.

Figure~\ref{fig:A4} (h) shows the LFs of the cluster and its tidal tails, with a KS statistic of 0.21 and a p-value of 0.13. Although the p-value indicates that the difference is not statistically significant, the KS statistic suggests a moderate divergence. The cluster retains a higher number of massive stars, whereas the tails are dominated by lower-mass stars. The binary fraction in the cluster was found to be $0.34 \pm 0.06$, compared to a lower average of $0.27$ in the tails. The leading and trailing tails show the binary fractions of $0.23 \pm 0.07$ and $0.31 \pm 0.09$, respectively.

\subsection{M67 (Collinder 204, MWSC 1585, Melotte 94, NGC 2682, OCL 549)}
\citet{2022A&A...659A..59T} reported halo stars but no tidal tails for M67, while \citet{2020PASJ...72...47G} identified 1618 likely members, including 85 extra-tidal stars beyond their tidal radius of 16.3~pc, and observed two tidal tails of $\approx$39~pc. In our analysis, we detected 1200 stars: 1037 cluster members, 163 tail stars, 110 extra-tidal stars (beyond the tidal radius of 16.3 pc), and 73 extra-tidal stars (beyond $r_J = 20.6$~pc). The leading and trailing tails span $\approx$29~pc and $\approx$31~pc, respectively, with a total tip-to-tip extent of $\approx$102~pc. This extent is slightly shorter than in some earlier reports, likely because previous estimates relied only on 2D sky projections. We account for the cluster's 3D orientation and proper motions to more conservatively identify tail members. This reduces contamination from foreground or background stars that may align by chance in projection, and avoids overestimating tail lengths based on stars not dynamically connected to the cluster. The orientation of the observed tails along the orbit, with tail stars at an angle, matches the theoretical expectations (Fig. \ref{fig:2}). Figure \ref{fig:3} reveals that in the cluster core (blue vectors), motion is predominantly oriented perpendicular to the orbit, lacking a clear rotational pattern. Tail vectors (black and red) show a slight counterclockwise motion, indicating weak rotation, with tail stars moving away from the expected orbital path.

Figure \ref{fig:A5} (h) presents the LFs for the cluster and its tidal tails, with a KS statistic of 0.10 and a p-value of 0.14, indicating no significant difference. A slight excess of low-mass stars in the tails suggests mild mass segregation, but overall, the stellar distributions remain comparable. The binary fraction in the cluster is found to be $0.28 \pm 0.02$, while the tails show a slightly higher average value of $0.30$. The leading and trailing tails exhibit the binary fractions of $0.33 \pm 0.09$ and $0.26 \pm 0.08$, respectively.

\section{Cluster rotation}
\label{sec:6}

Our analysis draws inspiration from the residual velocity method outlined in the study by \citet{2022ApJ...938..100H,2024ApJ...963..153H}, and the methodology proceeds as follows:
\begin{enumerate}
    \item We first confirm the presence of rotation within the cluster and its tail stars by analyzing the velocity distribution of stars.
    \item Next, we determine the angles (\( \alpha, \beta, \gamma \)) that represent the rotation axis in the Galactic Cartesian coordinate system.
    \item Finally, we compute the rotational velocity using the derived angles.
\end{enumerate}

To validate our method, we applied it to the Praesepe cluster using data from \citet{2022ApJ...938..100H} and compared the results with their reported values. Our derived mean azimuthal velocity ($v_{\varphi}$) is $0.1 \pm 0.07$ km s$^{-1}$ within the tidal radius of 10 pc and $0.2 \pm 0.07$ km s$^{-1}$ for the full sample, in close agreement with \citet{2022ApJ...938..100H,2024ApJ...963..153H}. The $v_{\varphi}$ and root mean squared (RMS) $v_{\varphi}$ profiles as a function of radius also showed consistency. Using their core radius estimates (\( r_{\mathrm{co}} = 1.6,\ 2.0,\ 2.6\,\mathrm{pc} \)), we computed tidal masses of \( M_t = (476 \pm 120),\ (521 \pm 132),\ (581 \pm 150)\,M_\odot \), respectively. These are also similar to the previously reported values of \( M_t = (537 \pm 146),\ (609 \pm 135),\ (\approx 510)\, M_\odot \), respectively, supporting the validity of our mass estimation method.

\subsection{Sample of the stars}
We identified cluster and tail stars with reliable RV values, which in our case are provided primarily for brighter stars with apparent magnitudes of \( G < 15 \) and stars with RV errors smaller than 5 km s$^{-1}$. In the case of M67, we obtained a sample of 458 stars with reliable RV measurements. The dataset for NGC 2281 was significantly smaller, containing only 123 stars with reliable RVs. Meanwhile, NGC 2354, Alessi 2, and BH 164 were excluded from the analysis, as their reliable RV data had very few stars. Such a small sample size was inadequate for establishing reliable statistical trends or drawing meaningful astrophysical conclusions about the clusters' kinematics.

\begin{figure*}[t]
\centering
\includegraphics[width=0.94\textwidth, height=0.43\textheight]{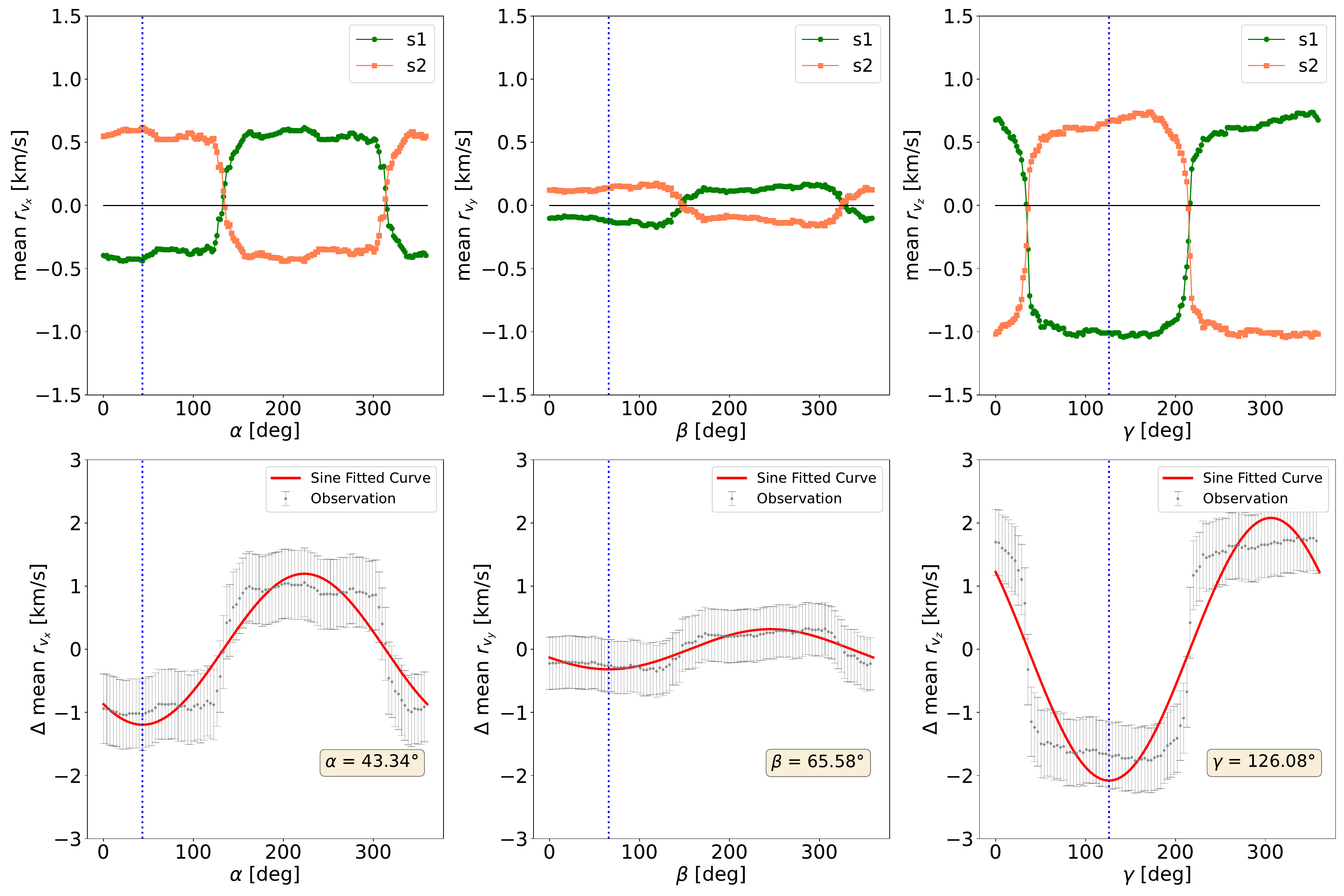} 
\caption{Mean residual velocities shown as a function of the three position angles ($\alpha$, $\beta$, $\gamma$) for the two subsamples in the top panel for M67 within 5$r_J$. Difference ($\Delta$) in the mean residual velocities ($v_{x_c}$, $v_{y_c}$, and $v_{z_c}$) of the two subsamples is displayed as a function of the position angles (PA) in the bottom panel. Error bars (in gray), best-fitting sine functions (in red), and the initially fitted angles are also shown.}
\label{fig:4}
\end{figure*}
\subsection{Coordinate systems and rotational analysis}
The analysis of cluster kinematics begins in the Galactic Cartesian coordinate system ($O_g$-$X_gY_gZ_g$), where the origin $O_g$ is at the Galactic center. 
We establish a cluster-centered coordinate system ($O_c$-$X_cY_cZ_c$) to analyze internal cluster dynamics by translating the origin to the cluster center. The transformation involves subtracting the cluster's central coordinates ($x_{g,s}$, $y_{g,s}$, $z_{g,s}$) and systemic velocities ($v_{x_{g,s}}$, $v_{y_{g,s}}$, $v_{z_{g,s}}$) from the coordinates and velocities of all the stars. This transformation yields the residual spatial coordinates ($X_c$,$Y_c$,$Z_c$ ) and the residual velocities ($v_{x_c}$, $v_{y_c}$, $v_{z_c}$) of individual stars relative to the cluster's systemic motion.

Following \citet{2022ApJ...938..100H,2024ApJ...963..153H}, we  determined the cluster's rotation axis $\overrightarrow{l}$ through three position angles ($\alpha$, $\beta$, $\gamma$). The determination process involves these key steps:
\begin{enumerate}
\item The projection of the rotation axis $\overrightarrow{l}$ onto the $Y_c$-$Z_c$ plane ($l_{yz}$) divides the cluster into two subsamples. For a rotating cluster, these subsamples should exhibit opposite signs in their mean residual velocities $v_{x_c}$.\\
\item Starting from $\alpha = 0^\circ$ (aligned with the $Y_c$-axis), we incrementally rotate $l_{yz}$ counterclockwise, computing the mean residual velocities $v_{x_c}$ for both subsamples at each angle.\\
\item The optimal angle $\alpha$ is identified where the difference in mean residual velocities between the subsamples reaches its extremum, indicating the true projection of $\overrightarrow{l}$ onto the $Y_c$-$Z_c$ plane.\\
\item The process is repeated to determine $\beta$ and $\gamma$, with these angles satisfying the relationship:
$\tan \alpha \tan \gamma = \tan \beta$.
\end{enumerate}
Upon determining the rotation axis, we established a rotational Cartesian coordinate system ($O_r$-$X_rY_rZ_r$) aligned with $\overrightarrow{l}$. The transformation between the cluster-centered and rotational coordinate systems is achieved through a direction cosine matrix, where the elements are defined by the included angles between the respective coordinate axes. The transformation equations incorporate these direction cosines to convert both spatial coordinates and velocity components. Finally, we convert to the cylindrical coordinate system ($r$, $\varphi$, $z$). This transformation provides a natural framework for studying rotation, where stellar positions and velocities are expressed in terms of radial ($v_r$), azimuthal ($v_{\varphi}$), and vertical ($v_z$) components relative to the rotation axis. 
\begin{figure}[t]
\centering
\includegraphics[width=0.43\textwidth, height=0.26\textheight]{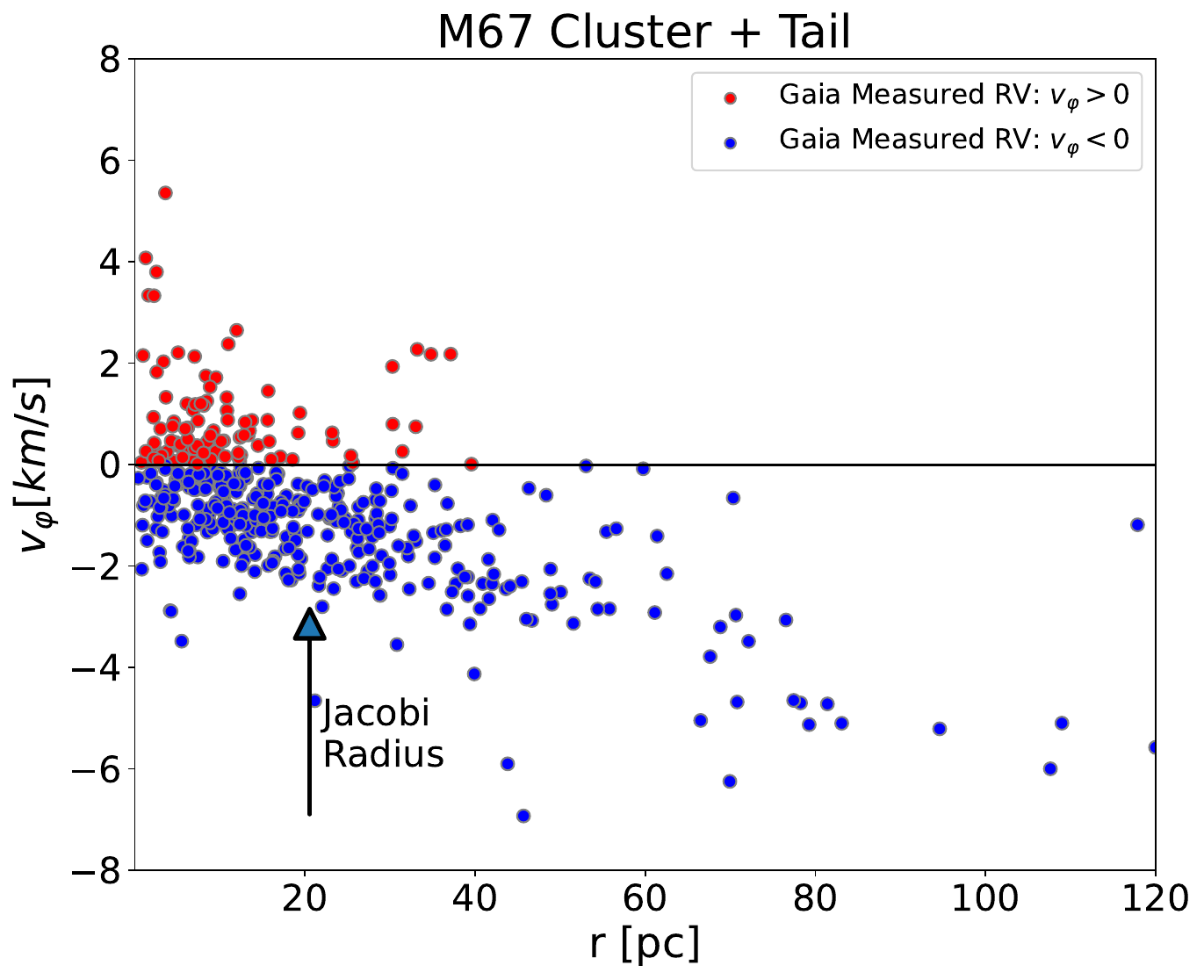}
\caption{Rotational velocity component as a function of the distance $r$ from the cluster center for the cluster + tail stars within 120 pc of M67. Similar plot for NGC 2281 in Fig. \ref{fig:A6}.}
\label{fig:5}
\end{figure}
\subsection{M67 rotation}
\begin{figure}
\centering
\includegraphics[width=0.43\textwidth, height=0.26\textheight]{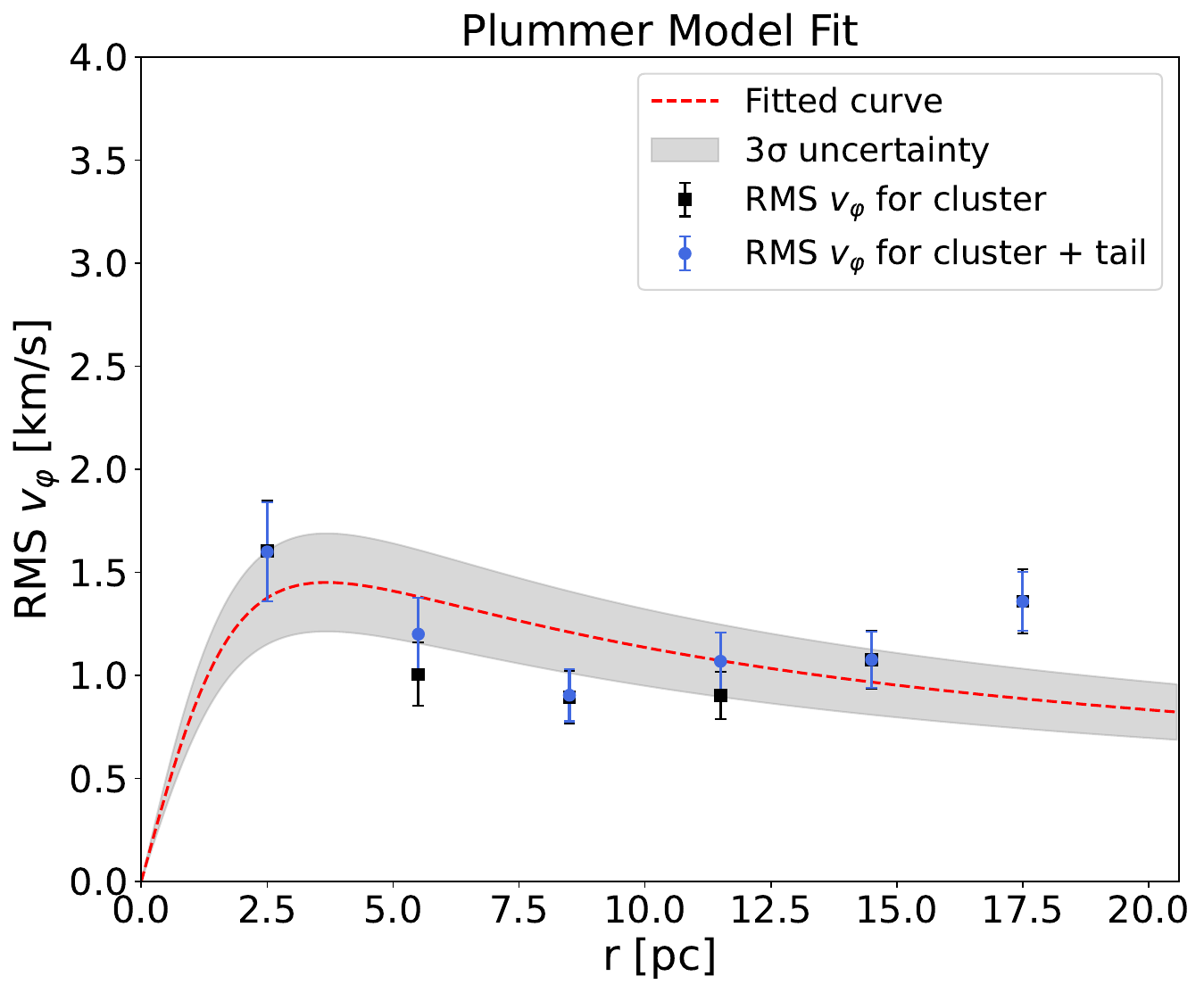}
\caption{Observed RMS \( v_\varphi \) of stars in M67 as a function of distance \(r\) from the cluster center (in blue and black) till 1${r_J}$. Gray shaded region shows the \(3\sigma\) uncertainty bounds. Red dotted line is the best-fitting Plummer model profile, obtained using a core radius of 2.6\,pc, fitted to the observed RMS \( v_\varphi \). Similar plot for NGC 2281 in Fig. \ref{fig:A6}.}
\label{fig:6}
\end{figure}

The mean residual velocities of the subsamples divided by the rotation axis for all the populations (cluster + tail, cluster-alone, and tails-alone) exhibited consistent opposite signs, indicating coherent rotation not only in the cluster core but also in the tidal tails, visible across all projection planes. Analysis of the RMS \( v_\varphi \) within the $r_J$  of M67 followed the expected trend from classical Newtonian dynamics, consistent with \citet{2022ApJ...938..100H,2024ApJ...963..153H}. Beyond 1\( r_J \), \( v_\varphi \) showed increasing dispersion (Fig. \ref{fig:5}) 
and a linear rise is seen in RMS \( v_\varphi \) beyond $\approx$15 pc (Fig. \ref{fig:6}), suggesting that tidal effects might be enhancing rotational motion in the outer regions of the cluster.

To determine the orientation of the rotation axis, we examined the residual velocities in the three planes within 5\( r_J \). The opposite signs in \( v_{x_c} \) and \( v_{z_c} \) in the \( Y_cZ_c \) and \( X_cY_c \) planes, respectively, confirmed a rotation signature, although low values of \( v_{y_c} \) in the \( X_cZ_c \) plane limited the reliability of \( \beta \). Sinusoidal fitting of mean residual velocity versus position angles (Fig. \ref{fig:4}) yielded position angles: \( \alpha = 43.3 \pm 3.6^\circ \), \( \beta = 65.6 \pm 10.0^\circ \), and \( \gamma = 126.1 \pm 2.0^\circ \). Using the trigonometric constraint: \( \tan \alpha \tan \gamma = \tan \beta \), we refined \( \beta \) to \( 127.7^\circ \). Using (\( \alpha, \beta, \gamma \)) = (\( 43.3^\circ, 127.7^\circ, 126.1^\circ \)), the mean \( v_\varphi \) was found to be \( -0.84 \pm 0.07\,\mathrm{km\,s^{-1}} \) for the entire cluster + tail system, \( -0.81 \pm 0.06\,\mathrm{km\,s^{-1}} \) for the cluster-alone, and \( -0.89 \pm 0.43\,\mathrm{km\,s^{-1}} \) for the tails-alone.

To study the rotational properties of the cluster in depth, we used the Plummer model \citep{10.1093/mnras/76.2.107} to fit the RMS $v_\varphi$ versus distance $r$ curve, which is as follows:

\begin{equation}
\rho(r) = \frac{3M_J}{4\pi r_{\mathrm{co}}^3} \left[1 + \left(\frac{r}{r_{\mathrm{co}}}\right)^2 \right]^{-5/2}.
\end{equation}

Here, $M_J$ is the total mass enclosed within the $r_J$ of the cluster, and $r_{\mathrm{co}}$ is the core radius. The corresponding enclosed mass $M(r)$ and circular speed $v_c(r)$ as functions of radius are expressed as

\begin{equation}
M(r) = M_J \frac{r^3}{\left(r^2 + r_{\mathrm{co}}^2\right)^{3/2}},
\end{equation}

\begin{equation}
v_c(r) = \left[ \frac{G M_t\ r^2}{\left(r^2 + r_{\mathrm{co}}^2\right)^{3/2}} \right]^{1/2}
\quad \Rightarrow \quad
v_c(r) = v_0 \frac{r}{\left(r^2 + r_{\mathrm{co}}^2\right)^{3/4}},
\end{equation}

where we defined the constant \( v_0 = \sqrt{G M_J} \). 
We fit the Plummer model to the observed rotation curve of M67 (Fig. \ref{fig:6}), using literature values for the core radius. \citet{2024A&A...686A..42H} reported \( r_{\mathrm{co}} = 2.6\,\mathrm{pc} \), while \citet{2022A&A...659A..59T} reported \( r_{\mathrm{co}} = 1.9\,\mathrm{pc} \). For \( r_{\mathrm{co}} = 2.6\,\mathrm{pc} \), we found a best-fit mass within the $r_J$ of \( M_J = (876 \pm 144)\,M_\odot \), whereas for \( r_{\mathrm{co}} = 1.9\,\mathrm{pc} \), the best-fit mass within the $r_J$ was \( M_J = (832 \pm 137)\,M_\odot \).
Integrated mass of M67 reported by \citet{2023MNRAS.525.2315A} is $(1843 \pm 368)\ M_\odot $ and mass within the $r_J$ reported by \citet{2024A&A...686A..42H} is $(2779 \pm 399)\ M_\odot $.

\subsection{NGC 2281 rotation}
\label{subsec:6.4}
The mean residual velocities of the subsamples divided by the rotation axis showed consistent opposite signs in the cluster + tail and cluster-alone populations, clearly visible only in the \( Y_cZ_c \) and \( X_cZ_c \) planes (Fig. \ref{fig:A6}). The \( X_cY_c \) plane showed minimal indication of rotation, so the angle $\gamma$ was unreliable in this case. The tails-alone population lacked a clear rotational signature and appeared to be dominated by noise. Within 1$r_J$ of NGC 2281, the RMS \( v_\varphi \) followed Newtonian expectations but showed significant dispersion (Fig. \ref{fig:A6}), which decreased beyond that Jacobi limit.

Sinusoidal fits to the differential mean residual velocities (\(v_{x_c}, v_{y_c}, v_{z_c}\)) versus the position angles (PAs) within 3\( r_J \) yielded characteristic rotation angles: \( \alpha = 124.9 \pm 9.1^\circ \), \( \beta = 65.6 \pm 5.1^\circ \), and \( \gamma = 109.8 \pm 14.0^\circ \). Using the tangent relation, we adopted (\( \alpha, \beta, \gamma \)) = (\(124.9^\circ\), \(65.6^\circ\), \(123.0^\circ\)), and computed the mean $ v_{\varphi}$ of $0.50 \pm 0.13\,\mathrm{km\,s^{-1}}$ and  $0.53 \pm 0.13\,\mathrm{km\,s^{-1}}$ for cluster + tail and cluster-alone populations, respectively, suggesting minimal rotational contribution from the tails.

We estimated the mass within the $r_J$ using the Plummer model fit to the RMS \( v_\varphi \) vs. distance r curve (Fig. \ref{fig:A6}), considering two core radii: \( r_{\mathrm{co}} = 2.6\,\mathrm{pc} \) \citep{2024A&A...686A..42H} and \( r_{\mathrm{co}} = 1.5\,\mathrm{pc} \) \citep{2022A&A...659A..59T}. The best-fit masses were \( M_J = (803 \pm 296)\,M_\odot \) for \(r_{\mathrm{co}} = 2.6\,\mathrm{pc} \) and \( (741 \pm 268)\,M_\odot \) for \( r_{\mathrm{co}} = 1.5\,\mathrm{pc} \). Integrated mass of NGC 2281 reported by \citet{2023MNRAS.525.2315A} is $(681 \pm 136)\ M_\odot$ and mass within the $r_J$ reported by  \citet{2024A&A...686A..42H} is $(885 \pm 112)\ M_\odot$.

\section{Discussion}
\label{sec:7}
We selected five open clusters with elongated morphology, as suggested by \citep{2022MNRAS.517.3525B, 2022A&A...659A..59T, Hunt2023A&A...673A.114H, 2024A&A...686A..42H, 2020PASJ...72...47G}, and used DBSCAN to identify their member stars. To detect tidal tails, we first examined specific properties indicative of tail structures, including RV and the CMD. We then applied an unsupervised machine learning approach, incorporating PCA to determine the elongation axis, followed by DBSCAN to identify the tails using Gaia DR3 data.
Our analysis revealed that all five clusters exhibit tail stars extending beyond their Jacobi limit (Fig. \ref{fig:2}). We corrected the proper motion vectors for projection effects (Subsect. \ref{subsec:correction}) and refined the spatial distribution of the tails by accounting for the spherical geometry of the sky (Subsect. \ref{subsec:tangential}). Figures \ref{fig:A1} to \ref{fig:A5} show detailed plots for each cluster, including their spatial distributions, CMDs, vector point diagrams, and LFs for both cluster members and tidal tails.

There is a notable increase in the number of stars identified in tidal tails across the clusters compared to what is documented in literature \citep{2022MNRAS.517.3525B, 2022A&A...659A..59T, Hunt2023A&A...673A.114H, 2024A&A...686A..42H, 2020PASJ...72...47G} with higher star count in four clusters (BH 164, Alessi 2, NGC 2281, and NGC 2354) and longer tidal tails in three (Alessi 2, NGC 2281, and NGC 2354). The improvements vary from case to case: star counts increased in our study by as few as 29 stars (Alessi 2) to as many as 131 (NGC 2281), as compared to the literature, while the tip-to-tip extent in our study increased by 8 pc (NGC 2354) up to 43 pc (Alessi 2) as compared to the literature. In M67, although we identified 1200 stars in total, the observed span of 102 pc was slightly shorter than that reported by \citet{2020PASJ...72...47G}. This difference may be due to our consideration of projection effects, which refines the estimation of tail lengths in the observed plane. Among the clusters studied, M67 exhibited the longest tidal tail span, followed by NGC 2281, Alessi 2, NGC 2354, and BH 164. The youngest cluster (BH 164, 65 Myr) showed the shortest tidal tail span, while the oldest (M67, 4 Gyr) exhibited the longest. According to N-body simulations \citep{2021A&A...647A.137J}, older clusters have longer tails; however, we did not detect complete tails ($\leq 40$ pc). Hence, the detected trend is a combination of the actual tail span, the different populations, and the incompleteness of the detection. 

In addition, clustering algorithms and statistical methods were used in this work to select stars with similar CMD locations and motions as tail candidates. However, such criteria can also be satisfied by field stars, leading to contamination of the sample, as seen in the clump we removed in NGC 2281 and in the unusually extended leading tail of NGC 2354 compared to simulations. Similar unexpected structures are also seen in other tail catalogs, for example, in the Pleiades (\citealt{Risbud2025arXiv250117225R}), which correspond to comoving stars but might not be real tails. More analysis is required to understand the completeness and purity of the tail candidates.

Proper motion vectors provided information about the motion of both the cluster and tail stars (Sect. \ref{sec:5}), shedding light on the overall dynamical state. These patterns indicated uniform alignment toward a specific direction, random disordered motion, or potential rotation. The LFs for all the clusters revealed that their tails notably lacked high-mass stars, which is expected behavior. When comparing the average binary fraction of tidal tails to that of cluster cores, we found that in three out of five clusters, the tails contained a higher proportion of binaries. Alessi 2 showed the most pronounced difference, with the tails averaging a binary fraction of 0.47 compared to 0.22 in the cluster core. NGC 2281 (tails: 0.28 vs. cluster: 0.24) and M67 (tails: 0.30 vs. cluster: 0.28) also followed this trend. BH 164 showed equal binary fractions (0.26) in both the tails and the cluster, while NGC 2354 was the only case where the tails had a lower average binary fraction (0.27) than the cluster (0.34). This generally higher binary fraction in tidal tails may result from dynamical interactions preferentially ejecting binary systems from cluster cores. All tidal tails demonstrated a clear deficiency of high-mass stars compared to their respective cluster centers. This deficiency aligns with theoretical predictions of mass segregation, where higher-mass stars sink toward cluster centers during dynamical evolution, while lower-mass stars and binaries are more susceptible to populating the outer regions and tidal tails. 

\citet{2023AJ....166..110P} observed that the binary fraction increases with radius in their study. \citet{2024A&A...691A.143W} used NBODY6 \citep{Aarseth1999PASP..111.1333A, Aarseth2003gnbs.book.....A} simulations to examine the binary population in clusters and tidal tails. They found that the binary fraction in the tidal tails fluctuates around the cluster's binary fraction with a steady decline in the binary fraction of the tails, which might lead to a lower binary fraction in the tail for very old clusters. However, more simulations and a better understanding of the incompleteness of the tidal tails are required to characterize the binary population in the tidal tails fully.

Based on simulations, the escaping stars from the tidal tails have preferred velocities where the leading tail stars move toward the Galactic center and the trailing tail stars move away from the Galactic center. This can be observed as an apparent rotation in the sky and the Galactic plane. We examined the clusters and their tidal tails using the residual velocity method to investigate their motion dynamics and 3D rotational behavior. 
We detected clear signatures of rotation in M67 and NGC 2281. However, BH 164, Alessi 2, and NGC 2354 were excluded from the rotation analysis due to insufficient reliable RV data. M67 demonstrated rotation not only in its central cluster but also extending into its tidal tails, with a mean \(v_\varphi\) of \(-0.84 \pm 0.07\) km s$^{-1}$ for the cluster and tail system. This rotational signature of M67, along with the expected sky positions, confirms that the extended structure around M67 is indeed its tidal tail. In contrast, NGC 2281 showed rotation primarily in its cluster region with \(v_\varphi = 0.53 \pm 0.13\) km s$^{-1}$, while its tails exhibited minimal contribution to the overall rotation. 

Both the clusters demonstrated rotation within 1$r_J$ that follows classical Newtonian dynamics, consistent with observations of other clusters such as Praesepe, Pleiades, Alpha Persei, and Hyades reported by \citet{2022ApJ...938..100H,2024ApJ...963..153H}. This suggested a fundamental similarity in the physical processes governing their core dynamics despite the differences in their extended structures. The mass estimations within the $r_J$ using Plummer model fits yielded the following results for the two clusters: M67’s ranges from \(832\)–\(876\,M_\odot\) and NGC 2281’s from \(741\)–\(803\,M_\odot\), depending on the assumed core radius values from literature.

\section{Summary}
\label{sec:8}
\begin{itemize}
    \item We used machine learning algorithms to identify core members and specific properties to detect the tidal tails in five clusters (BH 164, Alessi 2, NGC 2281, NGC 2354, and M67) using Gaia DR3 data. The clusters have ages of 60 Myr to 4 Gyr and are at distances of 400 to 1300 pc.
    \item The clusters showed tidal tails extending beyond their Jacobi limit, ranging from one to four times their $r_J$. 
    \item In BH 164, we identified 396 stars with two tidal tails forming an S-shaped structure, spanning approximately 40 pc. Alessi 2 contained 230 stars with a subtle \reflectbox{S}-like shape extending 90 pc. NGC 2281 hosted 574 stars, featuring a clear \reflectbox{S}-structure; its leading and trailing tails measured 39 pc and 25 pc, respectively, resulting in a total extent of about 91 pc. NGC 2354 contained 386 stars, with tails of 25 pc (leading) and 14 pc (trailing), extending 69 pc in total. Finally, M67 hosted 1200 stars, with a leading tail of 29 pc and a trailing tail of 31 pc, extending to a total of approximately 102 pc.
    
    \item The M67 cluster and its tidal tails displayed a sign of rotation in the sky plane, aligning with predictions from N-body simulations and confirmed through rotational analysis, indicating that the extended structure is indeed a tidal tail. While NGC 2281 also revealed rotational features after the rotational analysis, these were confined primarily to its central cluster regions. For the remaining clusters, the available RV data were insufficient to confidently assess the presence of rotation.
    \item In this study, the binary fraction and the LF of the tails and the main cluster are compared. In general, the binary fraction in the tidal tails is greater than in the cluster. Concerning the mass function, we observed that the overall slope remained consistent between clusters and tidal tails across all five clusters, but all tails showed a deficiency of high-mass stars.

\end{itemize} 
Our results showed that carefully optimized clustering methods can identify extended structures around star clusters up to a kiloparsec. This broadens the range of clusters that can be studied compared to methods relying solely on projected on-sky parameters (e.g., \citealt{2022MNRAS.517.3525B, 2022A&A...659A..59T}), probabilistic approaches (\citealt{2024A&A...691A..28K}), or techniques constrained by astrometric precision, such as the convergent point method, which is typically limited to within $\approx$400 pc (\citealt{Risbud2025arXiv250117225R}). Detecting rotational patterns in outer regions also serves as an important tool for validating the presence of tidal tails. Extending this analysis to a larger sample of clusters could reveal hidden tidal structures in the Galactic field and provide important observational constraints for theoretical analysis.

\section*{Data availability}
Table members.dat is only available in electronic form at the CDS via anonymous ftp to \href{https://cdsarc.u-strasbg.fr/}{cdsarc.u-strasbg.fr} (130.79.128.5) or via \href{https://cdsarc.cds.unistra.fr/viz-bin/qcat?J/A+A/}{http://cdsweb.u-strasbg.fr/cgi-bin/qcat?J/A+A/}.

\begin{acknowledgements}
We thank the anonymous referee for their constructive comments, which helped improve the manuscript. IS thanks Prof. Kulinder Pal Singh, who provided support during the initial stages of this research. VJ thanks the Alexander von Humboldt Foundation for their support. This study has used data from the European Space Agency (ESA) Gaia mission (\href{https://www.cosmos.esa.int/Gaia}{https://www.cosmos.esa.int/Gaia}), processed by the Gaia Data Processing and Analysis Consortium (DPAC, \href{https://www.cosmos.esa.int/web/Gaia/dpac/consortium}{https://www.cosmos.esa.int/web/Gaia/dpac/consortium}). Funding for the DPAC has been provided by national institutions, in particular the institutions participating in the Gaia Multilateral Agreement.
\end{acknowledgements}

\bibliographystyle{aa} 
\bibliography{references}

\begin{thebibliography}{67}
\expandafter\ifx\csname natexlab\endcsname\relax\def\natexlab#1{#1}\fi

\bibitem[{{Aarseth}(1999)}]{Aarseth1999PASP..111.1333A}
{Aarseth}, S.~J. 1999, \pasp, 111, 1333

\bibitem[{{Aarseth}(2003)}]{Aarseth2003gnbs.book.....A}
{Aarseth}, S.~J. 2003, {Gravitational N-Body Simulations}

\bibitem[{{Almeida} {et~al.}(2023){Almeida}, {Monteiro}, \&
  {Dias}}]{2023MNRAS.525.2315A}
{Almeida}, A., {Monteiro}, H., \& {Dias}, W.~S. 2023, \mnras, 525, 2315

\bibitem[{{Andrews} {et~al.}(2022){Andrews}, {Curtis}, {Chanam{\'e}},
  {Ag{\"u}eros}, {Schuler}, {Kounkel}, \& {Covey}}]{2022AJ....163..275A}
{Andrews}, J.~J., {Curtis}, J.~L., {Chanam{\'e}}, J., {et~al.} 2022, \aj, 163,
  275

\bibitem[{{Astropy Collaboration} {et~al.}(2022){Astropy Collaboration},
  {Price-Whelan}, {Lim}, {Earl}, {Starkman}, {Bradley}, {Shupe}, {Patil},
  {Corrales}, {Brasseur}, {N{\"o}the}, {Donath}, {Tollerud}, {Morris},
  {Ginsburg}, {Vaher}, {Weaver}, {Tocknell}, {Jamieson}, {van Kerkwijk},
  {Robitaille}, {Merry}, {Bachetti}, {G{\"u}nther}, {Aldcroft},
  {Alvarado-Montes}, {Archibald}, {B{\'o}di}, {Bapat}, {Barentsen},
  {Baz{\'a}n}, {Biswas}, {Boquien}, {Burke}, {Cara}, {Cara}, {Conroy},
  {Conseil}, {Craig}, {Cross}, {Cruz}, {D'Eugenio}, {Dencheva}, {Devillepoix},
  {Dietrich}, {Eigenbrot}, {Erben}, {Ferreira}, {Foreman-Mackey}, {Fox},
  {Freij}, {Garg}, {Geda}, {Glattly}, {Gondhalekar}, {Gordon}, {Grant},
  {Greenfield}, {Groener}, {Guest}, {Gurovich}, {Handberg}, {Hart},
  {Hatfield-Dodds}, {Homeier}, {Hosseinzadeh}, {Jenness}, {Jones}, {Joseph},
  {Kalmbach}, {Karamehmetoglu}, {Ka{\l}uszy{\'n}ski}, {Kelley}, {Kern},
  {Kerzendorf}, {Koch}, {Kulumani}, {Lee}, {Ly}, {Ma}, {MacBride}, {Maljaars},
  {Muna}, {Murphy}, {Norman}, {O'Steen}, {Oman}, {Pacifici}, {Pascual},
  {Pascual-Granado}, {Patil}, {Perren}, {Pickering}, {Rastogi}, {Roulston},
  {Ryan}, {Rykoff}, {Sabater}, {Sakurikar}, {Salgado}, {Sanghi}, {Saunders},
  {Savchenko}, {Schwardt}, {Seifert-Eckert}, {Shih}, {Jain}, {Shukla}, {Sick},
  {Simpson}, {Singanamalla}, {Singer}, {Singhal}, {Sinha}, {Sip{\H{o}}cz},
  {Spitler}, {Stansby}, {Streicher}, {{\v{S}}umak}, {Swinbank}, {Taranu},
  {Tewary}, {Tremblay}, {de Val-Borro}, {Van Kooten}, {Vasovi{\'c}}, {Verma},
  {de Miranda Cardoso}, {Williams}, {Wilson}, {Winkel}, {Wood-Vasey}, {Xue},
  {Yoachim}, {Zhang}, {Zonca}, \& {Astropy Project
  Contributors}}]{2022ApJ...935..167A}
{Astropy Collaboration}, {Price-Whelan}, A.~M., {Lim}, P.~L., {et~al.} 2022,
  \apj, 935, 167

\bibitem[{{Astropy Collaboration} {et~al.}(2018){Astropy Collaboration},
  {Price-Whelan}, {Sip{\H{o}}cz}, {G{\"u}nther}, {Lim}, {Crawford}, {Conseil},
  {Shupe}, {Craig}, {Dencheva}, {Ginsburg}, {VanderPlas}, {Bradley},
  {P{\'e}rez-Su{\'a}rez}, {de Val-Borro}, {Aldcroft}, {Cruz}, {Robitaille},
  {Tollerud}, {Ardelean}, {Babej}, {Bach}, {Bachetti}, {Bakanov}, {Bamford},
  {Barentsen}, {Barmby}, {Baumbach}, {Berry}, {Biscani}, {Boquien}, {Bostroem},
  {Bouma}, {Brammer}, {Bray}, {Breytenbach}, {Buddelmeijer}, {Burke},
  {Calderone}, {Cano Rodr{\'\i}guez}, {Cara}, {Cardoso}, {Cheedella}, {Copin},
  {Corrales}, {Crichton}, {D'Avella}, {Deil}, {Depagne}, {Dietrich}, {Donath},
  {Droettboom}, {Earl}, {Erben}, {Fabbro}, {Ferreira}, {Finethy}, {Fox},
  {Garrison}, {Gibbons}, {Goldstein}, {Gommers}, {Greco}, {Greenfield},
  {Groener}, {Grollier}, {Hagen}, {Hirst}, {Homeier}, {Horton}, {Hosseinzadeh},
  {Hu}, {Hunkeler}, {Ivezi{\'c}}, {Jain}, {Jenness}, {Kanarek}, {Kendrew},
  {Kern}, {Kerzendorf}, {Khvalko}, {King}, {Kirkby}, {Kulkarni}, {Kumar},
  {Lee}, {Lenz}, {Littlefair}, {Ma}, {Macleod}, {Mastropietro}, {McCully},
  {Montagnac}, {Morris}, {Mueller}, {Mumford}, {Muna}, {Murphy}, {Nelson},
  {Nguyen}, {Ninan}, {N{\"o}the}, {Ogaz}, {Oh}, {Parejko}, {Parley}, {Pascual},
  {Patil}, {Patil}, {Plunkett}, {Prochaska}, {Rastogi}, {Reddy Janga},
  {Sabater}, {Sakurikar}, {Seifert}, {Sherbert}, {Sherwood-Taylor}, {Shih},
  {Sick}, {Silbiger}, {Singanamalla}, {Singer}, {Sladen}, {Sooley},
  {Sornarajah}, {Streicher}, {Teuben}, {Thomas}, {Tremblay}, {Turner},
  {Terr{\'o}n}, {van Kerkwijk}, {de la Vega}, {Watkins}, {Weaver}, {Whitmore},
  {Woillez}, {Zabalza}, \& {Astropy Contributors}}]{2018AJ....156..123A}
{Astropy Collaboration}, {Price-Whelan}, A.~M., {Sip{\H{o}}cz}, B.~M., {et~al.}
  2018, \aj, 156, 123

\bibitem[{{Astropy Collaboration} {et~al.}(2013){Astropy Collaboration},
  {Robitaille}, {Tollerud}, {Greenfield}, {Droettboom}, {Bray}, {Aldcroft},
  {Davis}, {Ginsburg}, {Price-Whelan}, {Kerzendorf}, {Conley}, {Crighton},
  {Barbary}, {Muna}, {Ferguson}, {Grollier}, {Parikh}, {Nair}, {Unther},
  {Deil}, {Woillez}, {Conseil}, {Kramer}, {Turner}, {Singer}, {Fox}, {Weaver},
  {Zabalza}, {Edwards}, {Azalee Bostroem}, {Burke}, {Casey}, {Crawford},
  {Dencheva}, {Ely}, {Jenness}, {Labrie}, {Lim}, {Pierfederici}, {Pontzen},
  {Ptak}, {Refsdal}, {Servillat}, \& {Streicher}}]{2013A&A...558A..33A}
{Astropy Collaboration}, {Robitaille}, T.~P., {Tollerud}, E.~J., {et~al.} 2013,
  \aap, 558, A33

\bibitem[{{Bergond} {et~al.}(2001){Bergond}, {Leon}, \&
  {Guibert}}]{2001A&A...377..462B}
{Bergond}, G., {Leon}, S., \& {Guibert}, J. 2001, \aap, 377, 462

\bibitem[{{Bhattacharya} {et~al.}(2022){Bhattacharya}, {Rao}, {Agarwal},
  {Balan}, \& {Vaidya}}]{2022MNRAS.517.3525B}
{Bhattacharya}, S., {Rao}, K.~K., {Agarwal}, M., {Balan}, S., \& {Vaidya}, K.
  2022, \mnras, 517, 3525

\bibitem[{{Bossini} {et~al.}(2019){Bossini}, {Vallenari}, {Bragaglia},
  {Cantat-Gaudin}, {Sordo}, {Balaguer-N{\'u}{\~n}ez}, {Jordi}, {Moitinho},
  {Soubiran}, {Casamiquela}, {Carrera}, \& {Heiter}}]{2019A&A...623A.108B}
{Bossini}, D., {Vallenari}, A., {Bragaglia}, A., {et~al.} 2019, \aap, 623, A108

\bibitem[{{Bovy}(2015)}]{2015ApJS..216...29B}
{Bovy}, J. 2015, \apjs, 216, 29

\bibitem[{{Bressan} {et~al.}(2012){Bressan}, {Marigo}, {Girardi}, {Salasnich},
  {Dal Cero}, {Rubele}, \& {Nanni}}]{Bressan2012MNRAS.427..127B}
{Bressan}, A., {Marigo}, P., {Girardi}, L., {et~al.} 2012, \mnras, 427, 127

\bibitem[{Cantat-Gaudin(2022)}]{universe8020111}
Cantat-Gaudin, T. 2022, Universe, 8

\bibitem[{{Cantat-Gaudin} {et~al.}(2020){Cantat-Gaudin}, {Anders},
  {Castro-Ginard}, {Jordi}, {Romero-G{\'o}mez}, {Soubiran}, {Casamiquela},
  {Tarricq}, {Moitinho}, {Vallenari}, {Bragaglia}, {Krone-Martins}, \&
  {Kounkel}}]{2020A&A...640A...1C}
{Cantat-Gaudin}, T., {Anders}, F., {Castro-Ginard}, A., {et~al.} 2020, \aap,
  640, A1

\bibitem[{{Cantat-Gaudin} {et~al.}(2018){Cantat-Gaudin}, {Jordi}, {Vallenari},
  {Bragaglia}, {Balaguer-N{\'u}{\~n}ez}, {Soubiran}, {Bossini}, {Moitinho},
  {Castro-Ginard}, {Krone-Martins}, {Casamiquela}, {Sordo}, \&
  {Carrera}}]{2018A&A...618A..93C}
{Cantat-Gaudin}, T., {Jordi}, C., {Vallenari}, A., {et~al.} 2018, \aap, 618,
  A93

\bibitem[{{Chumak} {et~al.}(2010){Chumak}, {Platais}, {McLaughlin},
  {Rastorguev}, \& {Chumak}}]{2010MNRAS.402.1841C}
{Chumak}, Y.~O., {Platais}, I., {McLaughlin}, D.~E., {Rastorguev}, A.~S., \&
  {Chumak}, O.~V. 2010, \mnras, 402, 1841

\bibitem[{{Combes} {et~al.}(1999){Combes}, {Leon}, \&
  {Meylan}}]{1999A&A...352..149C}
{Combes}, F., {Leon}, S., \& {Meylan}, G. 1999, \aap, 352, 149

\bibitem[{{Deason} {et~al.}(2019){Deason}, {Belokurov}, \&
  {Sanders}}]{2019MNRAS.490.3426D}
{Deason}, A.~J., {Belokurov}, V., \& {Sanders}, J.~L. 2019, \mnras, 490, 3426

\bibitem[{Ester {et~al.}(1996)Ester, Kriegel, Sander, \&
  Xu}]{10.5555/3001460.3001507}
Ester, M., Kriegel, H.-P., Sander, J., \& Xu, X. 1996, in Proceedings of the
  Second International Conference on Knowledge Discovery and Data Mining,
  KDD'96 (AAAI Press), 226–231

\bibitem[{{Gaia Collaboration} {et~al.}(2018){Gaia Collaboration}, {Brown},
  {Vallenari}, {Prusti}, {de Bruijne}, {Babusiaux}, {Bailer-Jones}, {Biermann},
  {Evans}, {Eyer}, {Jansen}, {Jordi}, {Klioner}, {Lammers}, {Lindegren},
  {Luri}, {Mignard}, {Panem}, {Pourbaix}, {Randich}, {Sartoretti}, {Siddiqui},
  {Soubiran}, {van Leeuwen}, {Walton}, {Arenou}, {Bastian}, {Cropper},
  {Drimmel}, {Katz}, {Lattanzi}, {Bakker}, {Cacciari}, {Casta{\~n}eda},
  {Chaoul}, {Cheek}, {De Angeli}, {Fabricius}, {Guerra}, {Holl}, {Masana},
  {Messineo}, {Mowlavi}, {Nienartowicz}, {Panuzzo}, {Portell}, {Riello},
  {Seabroke}, {Tanga}, {Th{\'e}venin}, {Gracia-Abril}, {Comoretto},
  {Garcia-Reinaldos}, {Teyssier}, {Altmann}, {Andrae}, {Audard},
  {Bellas-Velidis}, {Benson}, {Berthier}, {Blomme}, {Burgess}, {Busso},
  {Carry}, {Cellino}, {Clementini}, {Clotet}, {Creevey}, {Davidson}, {De
  Ridder}, {Delchambre}, {Dell'Oro}, {Ducourant},
  {Fern{\'a}ndez-Hern{\'a}ndez}, {Fouesneau}, {Fr{\'e}mat}, {Galluccio},
  {Garc{\'\i}a-Torres}, {Gonz{\'a}lez-N{\'u}{\~n}ez}, {Gonz{\'a}lez-Vidal},
  {Gosset}, {Guy}, {Halbwachs}, {Hambly}, {Harrison}, {Hern{\'a}ndez},
  {Hestroffer}, {Hodgkin}, {Hutton}, {Jasniewicz}, {Jean-Antoine-Piccolo},
  {Jordan}, {Korn}, {Krone-Martins}, {Lanzafame}, {Lebzelter}, {L{\"o}ffler},
  {Manteiga}, {Marrese}, {Mart{\'\i}n-Fleitas}, {Moitinho}, {Mora}, {Muinonen},
  {Osinde}, {Pancino}, {Pauwels}, {Petit}, {Recio-Blanco}, {Richards},
  {Rimoldini}, {Robin}, {Sarro}, {Siopis}, {Smith}, {Sozzetti}, {S{\"u}veges},
  {Torra}, {van Reeven}, {Abbas}, {Abreu Aramburu}, {Accart}, {Aerts},
  {Altavilla}, {{\'A}lvarez}, {Alvarez}, {Alves}, {Anderson}, {Andrei},
  {Anglada Varela}, {Antiche}, {Antoja}, {Arcay}, {Astraatmadja}, {Bach},
  {Baker}, {Balaguer-N{\'u}{\~n}ez}, {Balm}, {Barache}, {Barata}, {Barbato},
  {Barblan}, {Barklem}, {Barrado}, {Barros}, {Barstow}, {Bartholom{\'e}
  Mu{\~n}oz}, {Bassilana}, {Becciani}, {Bellazzini}, {Berihuete}, {Bertone},
  {Bianchi}, {Bienaym{\'e}}, {Blanco-Cuaresma}, {Boch}, {Boeche}, {Bombrun},
  {Borrachero}, {Bossini}, {Bouquillon}, {Bourda}, {Bragaglia}, {Bramante},
  {Breddels}, {Bressan}, {Brouillet}, {Br{\"u}semeister}, {Brugaletta},
  {Bucciarelli}, {Burlacu}, {Busonero}, {Butkevich}, {Buzzi}, {Caffau},
  {Cancelliere}, {Cannizzaro}, {Cantat-Gaudin}, {Carballo}, {Carlucci},
  {Carrasco}, {Casamiquela}, {Castellani}, {Castro-Ginard}, {Charlot},
  {Chemin}, {Chiavassa}, {Cocozza}, {Costigan}, {Cowell}, {Crifo}, {Crosta},
  {Crowley}, {Cuypers}, {Dafonte}, {Damerdji}, {Dapergolas}, {David}, {David},
  {de Laverny}, {De Luise}, {De March}, {de Martino}, {de Souza}, {de Torres},
  {Debosscher}, {del Pozo}, {Delbo}, {Delgado}, {Delgado}, {Di Matteo},
  {Diakite}, {Diener}, {Distefano}, {Dolding}, {Drazinos}, {Dur{\'a}n},
  {Edvardsson}, {Enke}, {Eriksson}, {Esquej}, {Eynard Bontemps}, {Fabre},
  {Fabrizio}, {Faigler}, {Falc{\~a}o}, {Farr{\`a}s Casas}, {Federici},
  {Fedorets}, {Fernique}, {Figueras}, {Filippi}, {Findeisen}, {Fonti},
  {Fraile}, {Fraser}, {Fr{\'e}zouls}, {Gai}, {Galleti}, {Garabato},
  {Garc{\'\i}a-Sedano}, {Garofalo}, {Garralda}, {Gavel}, {Gavras}, {Gerssen},
  {Geyer}, {Giacobbe}, {Gilmore}, {Girona}, {Giuffrida}, {Glass}, {Gomes},
  {Granvik}, {Gueguen}, {Guerrier}, {Guiraud}, {Guti{\'e}rrez-S{\'a}nchez},
  {Haigron}, {Hatzidimitriou}, {Hauser}, {Haywood}, {Heiter}, {Helmi}, {Heu},
  {Hilger}, {Hobbs}, {Hofmann}, {Holland}, {Huckle}, {Hypki}, {Icardi},
  {Jan{\ss}en}, {Jevardat de Fombelle}, {Jonker}, {Juh{\'a}sz}, {Julbe},
  {Karampelas}, {Kewley}, {Klar}, {Kochoska}, {Kohley}, {Kolenberg},
  {Kontizas}, {Kontizas}, {Koposov}, {Kordopatis}, {Kostrzewa-Rutkowska},
  {Koubsky}, {Lambert}, {Lanza}, {Lasne}, {Lavigne}, {Le Fustec}, {Le
  Poncin-Lafitte}, {Lebreton}, {Leccia}, {Leclerc}, {Lecoeur-Taibi},
  {Lenhardt}, {Leroux}, {Liao}, {Licata}, {Lindstr{\o}m}, {Lister}, {Livanou},
  {Lobel}, {L{\'o}pez}, {Managau}, {Mann}, {Mantelet}, {Marchal}, {Marchant},
  {Marconi}, {Marinoni}, {Marschalk{\'o}}, {Marshall}, {Martino}, {Marton},
  {Mary}, {Massari}, {Matijevi{\v{c}}}, {Mazeh}, {McMillan}, {Messina},
  {Michalik}, {Millar}, {Molina}, {Molinaro}, {Moln{\'a}r}, {Montegriffo},
  {Mor}, {Morbidelli}, {Morel}, {Morris}, {Mulone}, {Muraveva}, {Musella},
  {Nelemans}, {Nicastro}, {Noval}, {O'Mullane}, {Ord{\'e}novic},
  {Ord{\'o}{\~n}ez-Blanco}, {Osborne}, {Pagani}, {Pagano}, {Pailler},
  {Palacin}, {Palaversa}, {Panahi}, {Pawlak}, {Piersimoni}, {Pineau}, {Plachy},
  {Plum}, {Poggio}, {Poujoulet}, {Pr{\v{s}}a}, {Pulone}, {Racero}, {Ragaini},
  {Rambaux}, {Ramos-Lerate}, {Regibo}, {Reyl{\'e}}, {Riclet}, {Ripepi}, {Riva},
  {Rivard}, {Rixon}, {Roegiers}, {Roelens}, {Romero-G{\'o}mez}, {Rowell},
  {Royer}, {Ruiz-Dern}, {Sadowski}, {Sagrist{\`a} Sell{\'e}s}, {Sahlmann},
  {Salgado}, {Salguero}, {Sanna}, {Santana-Ros}, {Sarasso}, {Savietto},
  {Schultheis}, {Sciacca}, {Segol}, {Segovia}, {S{\'e}gransan}, {Shih},
  {Siltala}, {Silva}, {Smart}, {Smith}, {Solano}, {Solitro}, {Sordo}, {Soria
  Nieto}, {Souchay}, {Spagna}, {Spoto}, {Stampa}, {Steele},
  {Steidelm{\"u}ller}, {Stephenson}, {Stoev}, {Suess}, {Surdej}, {Szabados},
  {Szegedi-Elek}, {Tapiador}, {Taris}, {Tauran}, {Taylor}, {Teixeira},
  {Terrett}, {Teyssandier}, {Thuillot}, {Titarenko}, {Torra Clotet}, {Turon},
  {Ulla}, {Utrilla}, {Uzzi}, {Vaillant}, {Valentini}, {Valette}, {van Elteren},
  {Van Hemelryck}, {van Leeuwen}, {Vaschetto}, {Vecchiato}, {Veljanoski},
  {Viala}, {Vicente}, {Vogt}, {von Essen}, {Voss}, {Votruba}, {Voutsinas},
  {Walmsley}, {Weiler}, {Wertz}, {Wevers}, {Wyrzykowski}, {Yoldas},
  {{\v{Z}}erjal}, {Ziaeepour}, {Zorec}, {Zschocke}, {Zucker}, {Zurbach}, \&
  {Zwitter}}]{2018A&A...616A...1G}
{Gaia Collaboration}, {Brown}, A.~G.~A., {Vallenari}, A., {et~al.} 2018, \aap,
  616, A1

\bibitem[{{Gaia Collaboration} {et~al.}(2021){Gaia Collaboration}, {Brown},
  {Vallenari}, {Prusti}, {de Bruijne}, {Babusiaux}, {Biermann}, {Creevey},
  {Evans}, {Eyer}, {Hutton}, {Jansen}, {Jordi}, {Klioner}, {Lammers},
  {Lindegren}, {Luri}, {Mignard}, {Panem}, {Pourbaix}, {Randich}, {Sartoretti},
  {Soubiran}, {Walton}, {Arenou}, {Bailer-Jones}, {Bastian}, {Cropper},
  {Drimmel}, {Katz}, {Lattanzi}, {van Leeuwen}, {Bakker}, {Cacciari},
  {Casta{\~n}eda}, {De Angeli}, {Ducourant}, {Fabricius}, {Fouesneau},
  {Fr{\'e}mat}, {Guerra}, {Guerrier}, {Guiraud}, {Jean-Antoine Piccolo},
  {Masana}, {Messineo}, {Mowlavi}, {Nicolas}, {Nienartowicz}, {Pailler},
  {Panuzzo}, {Riclet}, {Roux}, {Seabroke}, {Sordo}, {Tanga}, {Th{\'e}venin},
  {Gracia-Abril}, {Portell}, {Teyssier}, {Altmann}, {Andrae}, {Bellas-Velidis},
  {Benson}, {Berthier}, {Blomme}, {Brugaletta}, {Burgess}, {Busso}, {Carry},
  {Cellino}, {Cheek}, {Clementini}, {Damerdji}, {Davidson}, {Delchambre},
  {Dell'Oro}, {Fern{\'a}ndez-Hern{\'a}ndez}, {Galluccio}, {Garc{\'\i}a-Lario},
  {Garcia-Reinaldos}, {Gonz{\'a}lez-N{\'u}{\~n}ez}, {Gosset}, {Haigron},
  {Halbwachs}, {Hambly}, {Harrison}, {Hatzidimitriou}, {Heiter},
  {Hern{\'a}ndez}, {Hestroffer}, {Hodgkin}, {Holl}, {Jan{\ss}en}, {Jevardat de
  Fombelle}, {Jordan}, {Krone-Martins}, {Lanzafame}, {L{\"o}ffler}, {Lorca},
  {Manteiga}, {Marchal}, {Marrese}, {Moitinho}, {Mora}, {Muinonen}, {Osborne},
  {Pancino}, {Pauwels}, {Petit}, {Recio-Blanco}, {Richards}, {Riello},
  {Rimoldini}, {Robin}, {Roegiers}, {Rybizki}, {Sarro}, {Siopis}, {Smith},
  {Sozzetti}, {Ulla}, {Utrilla}, {van Leeuwen}, {van Reeven}, {Abbas}, {Abreu
  Aramburu}, {Accart}, {Aerts}, {Aguado}, {Ajaj}, {Altavilla}, {{\'A}lvarez},
  {{\'A}lvarez Cid-Fuentes}, {Alves}, {Anderson}, {Anglada Varela}, {Antoja},
  {Audard}, {Baines}, {Baker}, {Balaguer-N{\'u}{\~n}ez}, {Balbinot}, {Balog},
  {Barache}, {Barbato}, {Barros}, {Barstow}, {Bartolom{\'e}}, {Bassilana},
  {Bauchet}, {Baudesson-Stella}, {Becciani}, {Bellazzini}, {Bernet}, {Bertone},
  {Bianchi}, {Blanco-Cuaresma}, {Boch}, {Bombrun}, {Bossini}, {Bouquillon},
  {Bragaglia}, {Bramante}, {Breedt}, {Bressan}, {Brouillet}, {Bucciarelli},
  {Burlacu}, {Busonero}, {Butkevich}, {Buzzi}, {Caffau}, {Cancelliere},
  {C{\'a}novas}, {Cantat-Gaudin}, {Carballo}, {Carlucci}, {Carnerero},
  {Carrasco}, {Casamiquela}, {Castellani}, {Castro-Ginard}, {Castro Sampol},
  {Chaoul}, {Charlot}, {Chemin}, {Chiavassa}, {Cioni}, {Comoretto}, {Cooper},
  {Cornez}, {Cowell}, {Crifo}, {Crosta}, {Crowley}, {Dafonte}, {Dapergolas},
  {David}, {David}, {de Laverny}, {De Luise}, {De March}, {De Ridder}, {de
  Souza}, {de Teodoro}, {de Torres}, {del Peloso}, {del Pozo}, {Delbo},
  {Delgado}, {Delgado}, {Delisle}, {Di Matteo}, {Diakite}, {Diener},
  {Distefano}, {Dolding}, {Eappachen}, {Edvardsson}, {Enke}, {Esquej}, {Fabre},
  {Fabrizio}, {Faigler}, {Fedorets}, {Fernique}, {Fienga}, {Figueras},
  {Fouron}, {Fragkoudi}, {Fraile}, {Franke}, {Gai}, {Garabato},
  {Garcia-Gutierrez}, {Garc{\'\i}a-Torres}, {Garofalo}, {Gavras}, {Gerlach},
  {Geyer}, {Giacobbe}, {Gilmore}, {Girona}, {Giuffrida}, {Gomel}, {Gomez},
  {Gonzalez-Santamaria}, {Gonz{\'a}lez-Vidal}, {Granvik},
  {Guti{\'e}rrez-S{\'a}nchez}, {Guy}, {Hauser}, {Haywood}, {Helmi}, {Hidalgo},
  {Hilger}, {H{\l}adczuk}, {Hobbs}, {Holland}, {Huckle}, {Jasniewicz},
  {Jonker}, {Juaristi Campillo}, {Julbe}, {Karbevska}, {Kervella}, {Khanna},
  {Kochoska}, {Kontizas}, {Kordopatis}, {Korn}, {Kostrzewa-Rutkowska},
  {Kruszy{\'n}ska}, {Lambert}, {Lanza}, {Lasne}, {Le Campion}, {Le Fustec},
  {Lebreton}, {Lebzelter}, {Leccia}, {Leclerc}, {Lecoeur-Taibi}, {Liao},
  {Licata}, {Lindstr{\o}m}, {Lister}, {Livanou}, {Lobel}, {Madrero Pardo},
  {Managau}, {Mann}, {Marchant}, {Marconi}, {Marcos Santos}, {Marinoni},
  {Marocco}, {Marshall}, {Martin Polo}, {Mart{\'\i}n-Fleitas}, {Masip},
  {Massari}, {Mastrobuono-Battisti}, {Mazeh}, {McMillan}, {Messina},
  {Michalik}, {Millar}, {Mints}, {Molina}, {Molinaro}, {Moln{\'a}r},
  {Montegriffo}, {Mor}, {Morbidelli}, {Morel}, {Morris}, {Mulone}, {Munoz},
  {Muraveva}, {Murphy}, {Musella}, {Noval}, {Ord{\'e}novic}, {Orr{\`u}},
  {Osinde}, {Pagani}, {Pagano}, {Palaversa}, {Palicio}, {Panahi}, {Pawlak},
  {Pe{\~n}alosa Esteller}, {Penttil{\"a}}, {Piersimoni}, {Pineau}, {Plachy},
  {Plum}, {Poggio}, {Poretti}, {Poujoulet}, {Pr{\v{s}}a}, {Pulone}, {Racero},
  {Ragaini}, {Rainer}, {Raiteri}, {Rambaux}, {Ramos}, {Ramos-Lerate}, {Re
  Fiorentin}, {Regibo}, {Reyl{\'e}}, {Ripepi}, {Riva}, {Rixon}, {Robichon},
  {Robin}, {Roelens}, {Rohrbasser}, {Romero-G{\'o}mez}, {Rowell}, {Royer},
  {Rybicki}, {Sadowski}, {Sagrist{\`a} Sell{\'e}s}, {Sahlmann}, {Salgado},
  {Salguero}, {Samaras}, {Sanchez Gimenez}, {Sanna}, {Santove{\~n}a},
  {Sarasso}, {Schultheis}, {Sciacca}, {Segol}, {Segovia}, {S{\'e}gransan},
  {Semeux}, {Shahaf}, {Siddiqui}, {Siebert}, {Siltala}, {Slezak}, {Smart},
  {Solano}, {Solitro}, {Souami}, {Souchay}, {Spagna}, {Spoto}, {Steele},
  {Steidelm{\"u}ller}, {Stephenson}, {S{\"u}veges}, {Szabados}, {Szegedi-Elek},
  {Taris}, {Tauran}, {Taylor}, {Teixeira}, {Thuillot}, {Tonello}, {Torra},
  {Torra}, {Turon}, {Unger}, {Vaillant}, {van Dillen}, {Vanel}, {Vecchiato},
  {Viala}, {Vicente}, {Voutsinas}, {Weiler}, {Wevers}, {Wyrzykowski}, {Yoldas},
  {Yvard}, {Zhao}, {Zorec}, {Zucker}, {Zurbach}, \&
  {Zwitter}}]{2021A&A...649A...1G}
{Gaia Collaboration}, {Brown}, A.~G.~A., {Vallenari}, A., {et~al.} 2021, \aap,
  649, A1

\bibitem[{{Gaia Collaboration} {et~al.}(2023){Gaia Collaboration}, {Vallenari},
  {Brown}, {Prusti}, {de Bruijne}, {Arenou}, {Babusiaux}, {Biermann},
  {Creevey}, {Ducourant}, {Evans}, {Eyer}, {Guerra}, {Hutton}, {Jordi},
  {Klioner}, {Lammers}, {Lindegren}, {Luri}, {Mignard}, {Panem}, {Pourbaix},
  {Randich}, {Sartoretti}, {Soubiran}, {Tanga}, {Walton}, {Bailer-Jones},
  {Bastian}, {Drimmel}, {Jansen}, {Katz}, {Lattanzi}, {van Leeuwen}, {Bakker},
  {Cacciari}, {Casta{\~n}eda}, {De Angeli}, {Fabricius}, {Fouesneau},
  {Fr{\'e}mat}, {Galluccio}, {Guerrier}, {Heiter}, {Masana}, {Messineo},
  {Mowlavi}, {Nicolas}, {Nienartowicz}, {Pailler}, {Panuzzo}, {Riclet}, {Roux},
  {Seabroke}, {Sordo}, {Th{\'e}venin}, {Gracia-Abril}, {Portell}, {Teyssier},
  {Altmann}, {Andrae}, {Audard}, {Bellas-Velidis}, {Benson}, {Berthier},
  {Blomme}, {Burgess}, {Busonero}, {Busso}, {C{\'a}novas}, {Carry}, {Cellino},
  {Cheek}, {Clementini}, {Damerdji}, {Davidson}, {de Teodoro}, {Nu{\~n}ez
  Campos}, {Delchambre}, {Dell'Oro}, {Esquej}, {Fern{\'a}ndez-Hern{\'a}ndez},
  {Fraile}, {Garabato}, {Garc{\'\i}a-Lario}, {Gosset}, {Haigron}, {Halbwachs},
  {Hambly}, {Harrison}, {Hern{\'a}ndez}, {Hestroffer}, {Hodgkin}, {Holl},
  {Jan{\ss}en}, {Jevardat de Fombelle}, {Jordan}, {Krone-Martins}, {Lanzafame},
  {L{\"o}ffler}, {Marchal}, {Marrese}, {Moitinho}, {Muinonen}, {Osborne},
  {Pancino}, {Pauwels}, {Recio-Blanco}, {Reyl{\'e}}, {Riello}, {Rimoldini},
  {Roegiers}, {Rybizki}, {Sarro}, {Siopis}, {Smith}, {Sozzetti}, {Utrilla},
  {van Leeuwen}, {Abbas}, {{\'A}brah{\'a}m}, {Abreu Aramburu}, {Aerts},
  {Aguado}, {Ajaj}, {Aldea-Montero}, {Altavilla}, {{\'A}lvarez}, {Alves},
  {Anders}, {Anderson}, {Anglada Varela}, {Antoja}, {Baines}, {Baker},
  {Balaguer-N{\'u}{\~n}ez}, {Balbinot}, {Balog}, {Barache}, {Barbato},
  {Barros}, {Barstow}, {Bartolom{\'e}}, {Bassilana}, {Bauchet}, {Becciani},
  {Bellazzini}, {Berihuete}, {Bernet}, {Bertone}, {Bianchi}, {Binnenfeld},
  {Blanco-Cuaresma}, {Blazere}, {Boch}, {Bombrun}, {Bossini}, {Bouquillon},
  {Bragaglia}, {Bramante}, {Breedt}, {Bressan}, {Brouillet}, {Brugaletta},
  {Bucciarelli}, {Burlacu}, {Butkevich}, {Buzzi}, {Caffau}, {Cancelliere},
  {Cantat-Gaudin}, {Carballo}, {Carlucci}, {Carnerero}, {Carrasco},
  {Casamiquela}, {Castellani}, {Castro-Ginard}, {Chaoul}, {Charlot}, {Chemin},
  {Chiaramida}, {Chiavassa}, {Chornay}, {Comoretto}, {Contursi}, {Cooper},
  {Cornez}, {Cowell}, {Crifo}, {Cropper}, {Crosta}, {Crowley}, {Dafonte},
  {Dapergolas}, {David}, {David}, {de Laverny}, {De Luise}, {De March}, {De
  Ridder}, {de Souza}, {de Torres}, {del Peloso}, {del Pozo}, {Delbo},
  {Delgado}, {Delisle}, {Demouchy}, {Dharmawardena}, {Di Matteo}, {Diakite},
  {Diener}, {Distefano}, {Dolding}, {Edvardsson}, {Enke}, {Fabre}, {Fabrizio},
  {Faigler}, {Fedorets}, {Fernique}, {Fienga}, {Figueras}, {Fournier},
  {Fouron}, {Fragkoudi}, {Gai}, {Garcia-Gutierrez}, {Garcia-Reinaldos},
  {Garc{\'\i}a-Torres}, {Garofalo}, {Gavel}, {Gavras}, {Gerlach}, {Geyer},
  {Giacobbe}, {Gilmore}, {Girona}, {Giuffrida}, {Gomel}, {Gomez},
  {Gonz{\'a}lez-N{\'u}{\~n}ez}, {Gonz{\'a}lez-Santamar{\'\i}a},
  {Gonz{\'a}lez-Vidal}, {Granvik}, {Guillout}, {Guiraud},
  {Guti{\'e}rrez-S{\'a}nchez}, {Guy}, {Hatzidimitriou}, {Hauser}, {Haywood},
  {Helmer}, {Helmi}, {Sarmiento}, {Hidalgo}, {Hilger}, {H{\l}adczuk}, {Hobbs},
  {Holland}, {Huckle}, {Jardine}, {Jasniewicz}, {Jean-Antoine Piccolo},
  {Jim{\'e}nez-Arranz}, {Jorissen}, {Juaristi Campillo}, {Julbe}, {Karbevska},
  {Kervella}, {Khanna}, {Kontizas}, {Kordopatis}, {Korn}, {K{\'o}sp{\'a}l},
  {Kostrzewa-Rutkowska}, {Kruszy{\'n}ska}, {Kun}, {Laizeau}, {Lambert},
  {Lanza}, {Lasne}, {Le Campion}, {Lebreton}, {Lebzelter}, {Leccia}, {Leclerc},
  {Lecoeur-Taibi}, {Liao}, {Licata}, {Lindstr{\o}m}, {Lister}, {Livanou},
  {Lobel}, {Lorca}, {Loup}, {Madrero Pardo}, {Magdaleno Romeo}, {Managau},
  {Mann}, {Manteiga}, {Marchant}, {Marconi}, {Marcos}, {Marcos Santos},
  {Mar{\'\i}n Pina}, {Marinoni}, {Marocco}, {Marshall}, {Martin Polo},
  {Mart{\'\i}n-Fleitas}, {Marton}, {Mary}, {Masip}, {Massari},
  {Mastrobuono-Battisti}, {Mazeh}, {McMillan}, {Messina}, {Michalik}, {Millar},
  {Mints}, {Molina}, {Molinaro}, {Moln{\'a}r}, {Monari}, {Mongui{\'o}},
  {Montegriffo}, {Montero}, {Mor}, {Mora}, {Morbidelli}, {Morel}, {Morris},
  {Muraveva}, {Murphy}, {Musella}, {Nagy}, {Noval}, {Oca{\~n}a}, {Ogden},
  {Ordenovic}, {Osinde}, {Pagani}, {Pagano}, {Palaversa}, {Palicio},
  {Pallas-Quintela}, {Panahi}, {Payne-Wardenaar}, {Pe{\~n}alosa Esteller},
  {Penttil{\"a}}, {Pichon}, {Piersimoni}, {Pineau}, {Plachy}, {Plum}, {Poggio},
  {Pr{\v{s}}a}, {Pulone}, {Racero}, {Ragaini}, {Rainer}, {Raiteri}, {Rambaux},
  {Ramos}, {Ramos-Lerate}, {Re Fiorentin}, {Regibo}, {Richards}, {Rios Diaz},
  {Ripepi}, {Riva}, {Rix}, {Rixon}, {Robichon}, {Robin}, {Robin}, {Roelens},
  {Rogues}, {Rohrbasser}, {Romero-G{\'o}mez}, {Rowell}, {Royer}, {Ruz Mieres},
  {Rybicki}, {Sadowski}, {S{\'a}ez N{\'u}{\~n}ez}, {Sagrist{\`a} Sell{\'e}s},
  {Sahlmann}, {Salguero}, {Samaras}, {Sanchez Gimenez}, {Sanna},
  {Santove{\~n}a}, {Sarasso}, {Schultheis}, {Sciacca}, {Segol}, {Segovia},
  {S{\'e}gransan}, {Semeux}, {Shahaf}, {Siddiqui}, {Siebert}, {Siltala},
  {Silvelo}, {Slezak}, {Slezak}, {Smart}, {Snaith}, {Solano}, {Solitro},
  {Souami}, {Souchay}, {Spagna}, {Spina}, {Spoto}, {Steele},
  {Steidelm{\"u}ller}, {Stephenson}, {S{\"u}veges}, {Surdej}, {Szabados},
  {Szegedi-Elek}, {Taris}, {Taylor}, {Teixeira}, {Tolomei}, {Tonello}, {Torra},
  {Torra}, {Torralba Elipe}, {Trabucchi}, {Tsounis}, {Turon}, {Ulla}, {Unger},
  {Vaillant}, {van Dillen}, {van Reeven}, {Vanel}, {Vecchiato}, {Viala},
  {Vicente}, {Voutsinas}, {Weiler}, {Wevers}, {Wyrzykowski}, {Yoldas}, {Yvard},
  {Zhao}, {Zorec}, {Zucker}, \& {Zwitter}}]{2023A&A...674A...1G}
{Gaia Collaboration}, {Vallenari}, A., {Brown}, A.~G.~A., {et~al.} 2023, \aap,
  674, A1

\bibitem[{{Gao}(2020{\natexlab{a}})}]{2020PASJ...72...47G}
{Gao}, X. 2020{\natexlab{a}}, \pasj, 72, 47

\bibitem[{{Gao}(2020{\natexlab{b}})}]{2020ApJ...894...48G}
{Gao}, X. 2020{\natexlab{b}}, \apj, 894, 48

\bibitem[{{Hao} {et~al.}(2022){Hao}, {Xu}, {Bian}, {Hou}, {Lin}, {Li}, \&
  {Liu}}]{2022ApJ...938..100H}
{Hao}, C.~J., {Xu}, Y., {Bian}, S.~B., {et~al.} 2022, \apj, 938, 100

\bibitem[{{Hao} {et~al.}(2024){Hao}, {Xu}, {Hou}, {Bian}, {Lin}, {Li}, {Dong},
  \& {Liu}}]{2024ApJ...963..153H}
{Hao}, C.~J., {Xu}, Y., {Hou}, L.~G., {et~al.} 2024, \apj, 963, 153

\bibitem[{{Helmi}(2020)}]{2020ARA&A..58..205H}
{Helmi}, A. 2020, \araa, 58, 205

\bibitem[{{Hou} \& {Han}(2014)}]{2014A&A...569A.125H}
{Hou}, L.~G. \& {Han}, J.~L. 2014, \aap, 569, A125

\bibitem[{{Hunt} \& {Reffert}(2023)}]{Hunt2023A&A...673A.114H}
{Hunt}, E.~L. \& {Reffert}, S. 2023, \aap, 673, A114

\bibitem[{{Hunt} \& {Reffert}(2024)}]{2024A&A...686A..42H}
{Hunt}, E.~L. \& {Reffert}, S. 2024, \aap, 686, A42

\bibitem[{{Hurley} {et~al.}(2013){Hurley}, {Tout}, \&
  {Pols}}]{2013ascl.soft03014H}
{Hurley}, J.~R., {Tout}, C.~A., \& {Pols}, O.~R. 2013, {BSE: Binary Star
  Evolution}, Astrophysics Source Code Library, record ascl:1303.014

\bibitem[{{Ibata} {et~al.}(2002){Ibata}, {Lewis}, {Irwin}, \&
  {Quinn}}]{2002MNRAS.332..915I}
{Ibata}, R.~A., {Lewis}, G.~F., {Irwin}, M.~J., \& {Quinn}, T. 2002, \mnras,
  332, 915

\bibitem[{{Jadhav} {et~al.}(2024){Jadhav}, {Kroupa}, {Wu}, {Pflamm-Altenburg},
  \& {Thies}}]{2024A&A...687A..89J}
{Jadhav}, V.~V., {Kroupa}, P., {Wu}, W., {Pflamm-Altenburg}, J., \& {Thies}, I.
  2024, \aap, 687, A89

\bibitem[{{Jadhav} {et~al.}(2025){Jadhav}, {Risbud}, {Kroupa}, \&
  {Wu}}]{Jadhav2025_tidal_tail_ii}
{Jadhav}, V.~V., {Risbud}, D., {Kroupa}, P., \& {Wu}, W. 2025, arXiv e-prints,
  arXiv:2508.15056

\bibitem[{{Jadhav} {et~al.}(2021){Jadhav}, {Roy}, {Joshi}, \&
  {Subramaniam}}]{Jadhav2021AJ....162..264J}
{Jadhav}, V.~V., {Roy}, K., {Joshi}, N., \& {Subramaniam}, A. 2021, \aj, 162,
  264

\bibitem[{{Jerabkova} {et~al.}(2021){Jerabkova}, {Boffin}, {Beccari}, {de
  Marchi}, {de Bruijne}, \& {Prusti}}]{2021A&A...647A.137J}
{Jerabkova}, T., {Boffin}, H. M.~J., {Beccari}, G., {et~al.} 2021, \aap, 647,
  A137

\bibitem[{{Kharchenko} {et~al.}(2012){Kharchenko}, {Piskunov}, {Schilbach},
  {R{\"o}ser}, \& {Scholz}}]{2012A&A...543A.156K}
{Kharchenko}, N.~V., {Piskunov}, A.~E., {Schilbach}, E., {R{\"o}ser}, S., \&
  {Scholz}, R.~D. 2012, \aap, 543, A156

\bibitem[{{Kharchenko} {et~al.}(2013){Kharchenko}, {Piskunov}, {Schilbach},
  {R{\"o}ser}, \& {Scholz}}]{2013A&A...558A..53K}
{Kharchenko}, N.~V., {Piskunov}, A.~E., {Schilbach}, E., {R{\"o}ser}, S., \&
  {Scholz}, R.~D. 2013, \aap, 558, A53

\bibitem[{{Kos}(2024)}]{2024A&A...691A..28K}
{Kos}, J. 2024, \aap, 691, A28

\bibitem[{{Kounkel} \& {Covey}(2019)}]{2019AJ....158..122K}
{Kounkel}, M. \& {Covey}, K. 2019, \aj, 158, 122

\bibitem[{{Kounkel} {et~al.}(2020){Kounkel}, {Covey}, \&
  {Stassun}}]{2020AJ....160..279K}
{Kounkel}, M., {Covey}, K., \& {Stassun}, K.~G. 2020, \aj, 160, 279

\bibitem[{{Kroupa}(2001)}]{2001MNRAS.322..231K}
{Kroupa}, P. 2001, \mnras, 322, 231

\bibitem[{{K{\"u}pper} {et~al.}(2011){K{\"u}pper}, {Maschberger}, {Kroupa}, \&
  {Baumgardt}}]{2011MNRAS.417.2300K}
{K{\"u}pper}, A. H.~W., {Maschberger}, T., {Kroupa}, P., \& {Baumgardt}, H.
  2011, \mnras, 417, 2300

\bibitem[{{Leon} {et~al.}(2000){Leon}, {Meylan}, \&
  {Combes}}]{2000A&A...359..907L}
{Leon}, S., {Meylan}, G., \& {Combes}, F. 2000, \aap, 359, 907

\bibitem[{{Li} {et~al.}(2020){Li}, {Shao}, {Li}, {Yu}, {Zhong}, \&
  {Chen}}]{Li2020ApJ...901...49L}
{Li}, L., {Shao}, Z., {Li}, Z.-Z., {et~al.} 2020, \apj, 901, 49

\bibitem[{{Li} {et~al.}(2021){Li}, {Li}, \& {Shao}}]{Li2021A&C....3600483L}
{Li}, Z.-Z., {Li}, L., \& {Shao}, Z. 2021, Astronomy and Computing, 36, 100483

\bibitem[{{Marks} \& {Kroupa}(2012)}]{2012A&A...543A...8M}
{Marks}, M. \& {Kroupa}, P. 2012, \aap, 543, A8

\bibitem[{{Meingast} {et~al.}(2019){Meingast}, {Alves}, \&
  {F{\"u}rnkranz}}]{2019A&A...622L..13M}
{Meingast}, S., {Alves}, J., \& {F{\"u}rnkranz}, V. 2019, \aap, 622, L13

\bibitem[{{Meingast} {et~al.}(2021){Meingast}, {Alves}, \&
  {Rottensteiner}}]{2021A&A...645A..84M}
{Meingast}, S., {Alves}, J., \& {Rottensteiner}, A. 2021, \aap, 645, A84

\bibitem[{Montuori {et~al.}(2007)Montuori, Capuzzo-Dolcetta, Matteo, Lepinette,
  \& Miocchi}]{Montuori_2007}
Montuori, M., Capuzzo-Dolcetta, R., Matteo, P.~D., Lepinette, A., \& Miocchi,
  P. 2007, The Astrophysical Journal, 659, 1212

\bibitem[{Odenkirchen {et~al.}(2003)Odenkirchen, Grebel, Dehnen, Rix, Yanny,
  Newberg, Rockosi, Martínez-Delgado, Brinkmann, \& Pier}]{Odenkirchen_2003}
Odenkirchen, M., Grebel, E.~K., Dehnen, W., {et~al.} 2003, The Astronomical
  Journal, 126, 2385

\bibitem[{{Odenkirchen} {et~al.}(2001){Odenkirchen}, {Grebel}, {Rockosi},
  {Dehnen}, {Ibata}, {Rix}, {Stolte}, {Wolf}, {Anderson}, {Bahcall},
  {Brinkmann}, {Csabai}, {Hennessy}, {Hindsley}, {Ivezi{\'c}}, {Lupton},
  {Munn}, {Pier}, {Stoughton}, \& {York}}]{2001ApJ...548L.165O}
{Odenkirchen}, M., {Grebel}, E.~K., {Rockosi}, C.~M., {et~al.} 2001, \apjl,
  548, L165

\bibitem[{{Pang} {et~al.}(2023){Pang}, {Wang}, {Tang}, {Rui}, {Bai}, {Li},
  {Feng}, {Kouwenhoven}, {Chen}, \& {Chuang}}]{2023AJ....166..110P}
{Pang}, X., {Wang}, Y., {Tang}, S.-Y., {et~al.} 2023, \aj, 166, 110

\bibitem[{{Pedregosa} {et~al.}(2011){Pedregosa}, {Varoquaux}, {Gramfort},
  {Michel}, {Thirion}, {Grisel}, {Blondel}, {M{\"u}ller}, {Nothman}, {Louppe},
  {Prettenhofer}, {Weiss}, {Dubourg}, {Vanderplas}, {Passos}, {Cournapeau},
  {Brucher}, {Perrot}, \& {Duchesnay}}]{2011JMLR...12.2825P}
{Pedregosa}, F., {Varoquaux}, G., {Gramfort}, A., {et~al.} 2011, Journal of
  Machine Learning Research, 12, 2825

\bibitem[{Plummer(1915)}]{10.1093/mnras/76.2.107}
Plummer, H.~C. 1915, Monthly Notices of the Royal Astronomical Society, 76, 107

\bibitem[{{Posti} \& {Helmi}(2019)}]{2019A&A...621A..56P}
{Posti}, L. \& {Helmi}, A. 2019, \aap, 621, A56

\bibitem[{{Ratzenb{\"o}ck} {et~al.}(2020){Ratzenb{\"o}ck}, {Meingast}, {Alves},
  {M{\"o}ller}, \& {Bomze}}]{2020A&A...639A..64R}
{Ratzenb{\"o}ck}, S., {Meingast}, S., {Alves}, J., {M{\"o}ller}, T., \&
  {Bomze}, I. 2020, \aap, 639, A64

\bibitem[{{Risbud} {et~al.}(2025){Risbud}, {Jadhav}, \&
  {Kroupa}}]{Risbud2025arXiv250117225R}
{Risbud}, D., {Jadhav}, V.~V., \& {Kroupa}, P. 2025, \aap, 694, A258

\bibitem[{{R{\"o}ser} \& {Schilbach}(2020)}]{2020A&A...638A...9R}
{R{\"o}ser}, S. \& {Schilbach}, E. 2020, \aap, 638, A9

\bibitem[{{R{\"o}ser} {et~al.}(2011){R{\"o}ser}, {Schilbach}, {Piskunov},
  {Kharchenko}, \& {Scholz}}]{2011A&A...531A..92R}
{R{\"o}ser}, S., {Schilbach}, E., {Piskunov}, A.~E., {Kharchenko}, N.~V., \&
  {Scholz}, R.~D. 2011, \aap, 531, A92

\bibitem[{Shen \& Zheng(2020)}]{Shen_2020}
Shen, J. \& Zheng, X.-W. 2020, Research in Astronomy and Astrophysics, 20, 159

\bibitem[{{Tarricq} {et~al.}(2022){Tarricq}, {Soubiran}, {Casamiquela},
  {Castro-Ginard}, {Olivares}, {Miret-Roig}, \& {Galli}}]{2022A&A...659A..59T}
{Tarricq}, Y., {Soubiran}, C., {Casamiquela}, L., {et~al.} 2022, \aap, 659, A59

\bibitem[{{van Leeuwen}(2009)}]{2009A&A...497..209V}
{van Leeuwen}, F. 2009, \aap, 497, 209

\bibitem[{{Vasiliev}(2019)}]{2019MNRAS.484.2832V}
{Vasiliev}, E. 2019, \mnras, 484, 2832

\bibitem[{{Wang} {et~al.}(2020){Wang}, {Iwasawa}, {Nitadori}, \&
  {Makino}}]{2020MNRAS.497..536W}
{Wang}, L., {Iwasawa}, M., {Nitadori}, K., \& {Makino}, J. 2020, \mnras, 497,
  536

\bibitem[{{Wirth} {et~al.}(2024){Wirth}, {Dinnbier}, {Kroupa}, \&
  {{\v{S}}ubr}}]{2024A&A...691A.143W}
{Wirth}, H., {Dinnbier}, F., {Kroupa}, P., \& {{\v{S}}ubr}, L. 2024, \aap, 691,
  A143

\bibitem[{{Zucker} {et~al.}(2022){Zucker}, {Peek}, \&
  {Loebman}}]{2022ApJ...936..160Z}
{Zucker}, C., {Peek}, J.~E.~G., \& {Loebman}, S. 2022, \apj, 936, 160

\end{thebibliography}
\begin{appendix}
\section{Supplementary query and figures}
\subsection{ADQL query for selecting the Gaia DR3 data}
\hrule

\begin{lstlisting}
SELECT * 
FROM gaiadr3.gaia_source
WHERE 
    1 = CONTAINS(
        POINT('ICRS', gaiadr3.gaia_source.ra, gaiadr3.gaia_source.dec),
        CIRCLE('ICRS', {ra}, {dec}, {radius})
    )
    AND gaiadr3.gaia_source.parallax > 0
    AND ABS(gaiadr3.gaia_source.pmra_error) < 0.5
    AND ABS(gaiadr3.gaia_source.pmdec_error) < 0.5
    AND ABS(gaiadr3.gaia_source.parallax_error / gaiadr3.gaia_source.parallax) < 0.1
    AND gaiadr3.gaia_source.pmra BETWEEN {pmra_min} AND {pmra_max}
    AND gaiadr3.gaia_source.pmdec BETWEEN {pmdec_min} AND {pmdec_max}  
\end{lstlisting}

\hrule
\vspace{3pt}
\begin{table}[h]
\centering
\small
\caption{Values for each cluster used in the ADQL query}
\begin{tabular}{lcccccc}
\toprule
Name &ra& dec & pmra (min) & pmra (max) & pmdec (min) & pmdec (max) \\
     & (deg) & (deg) & (mas yr$^{-1}$) & (mas yr$^{-1}$) & (mas yr$^{-1}$) & (mas yr$^{-1}$) \\
\midrule
BH 164    & 223.4  & -66.5 & -9.0  & -6.0  & -12.0 & -9.5 \\
Alessi 2  & 71.1   & 54.7  & -2.0  & -0.3  & -1.4  & -0.5 \\
NGC 2281  & 101.9  & 40.3  & -4.0  & -2.0  & -9.5  & -7.0 \\
NGC 2354  & 108.7  & -25.6 & -3.3  & -2.3  & 1.4   & 2.4  \\
M67       & 132.8  & 11.8  & -12.2 & -9.8  & -3.7  & -2.0 \\
\bottomrule
\end{tabular}
\label{tab:cluster_ra_dec_pm}
\end{table}

\subsection{Appendix figures of individual clusters and
their description} 
\label{subsec:appendix A2}
\begin{figure*}
\centering
\includegraphics[width=1\textwidth]{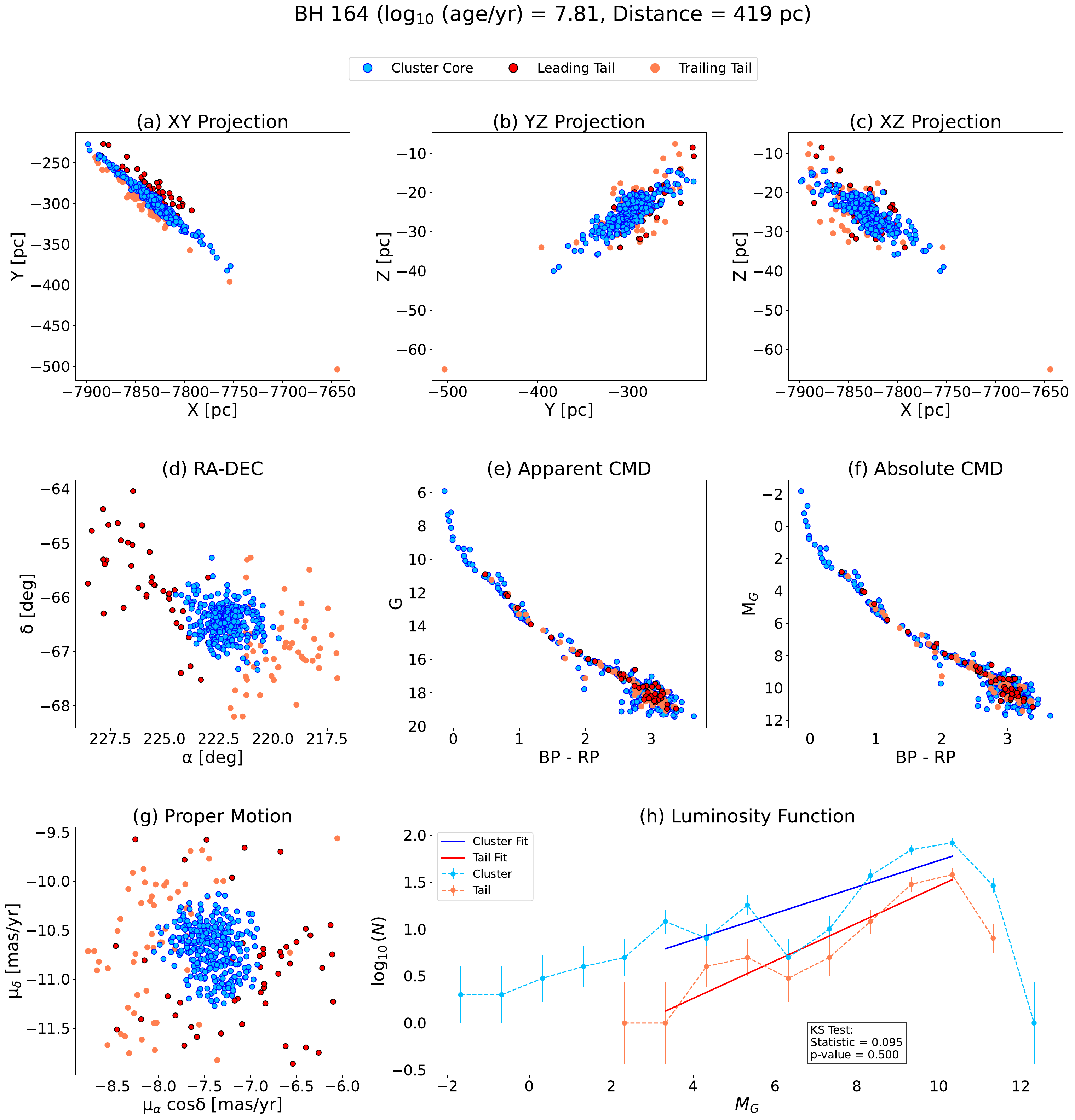} 
\caption{Results from our analysis for BH 164: Image description can be found in the Appendix \ref{subsec:appendix A2}.}
\label{fig:A1}
\end{figure*}

\begin{figure*}
\centering
\includegraphics[width=1\textwidth]{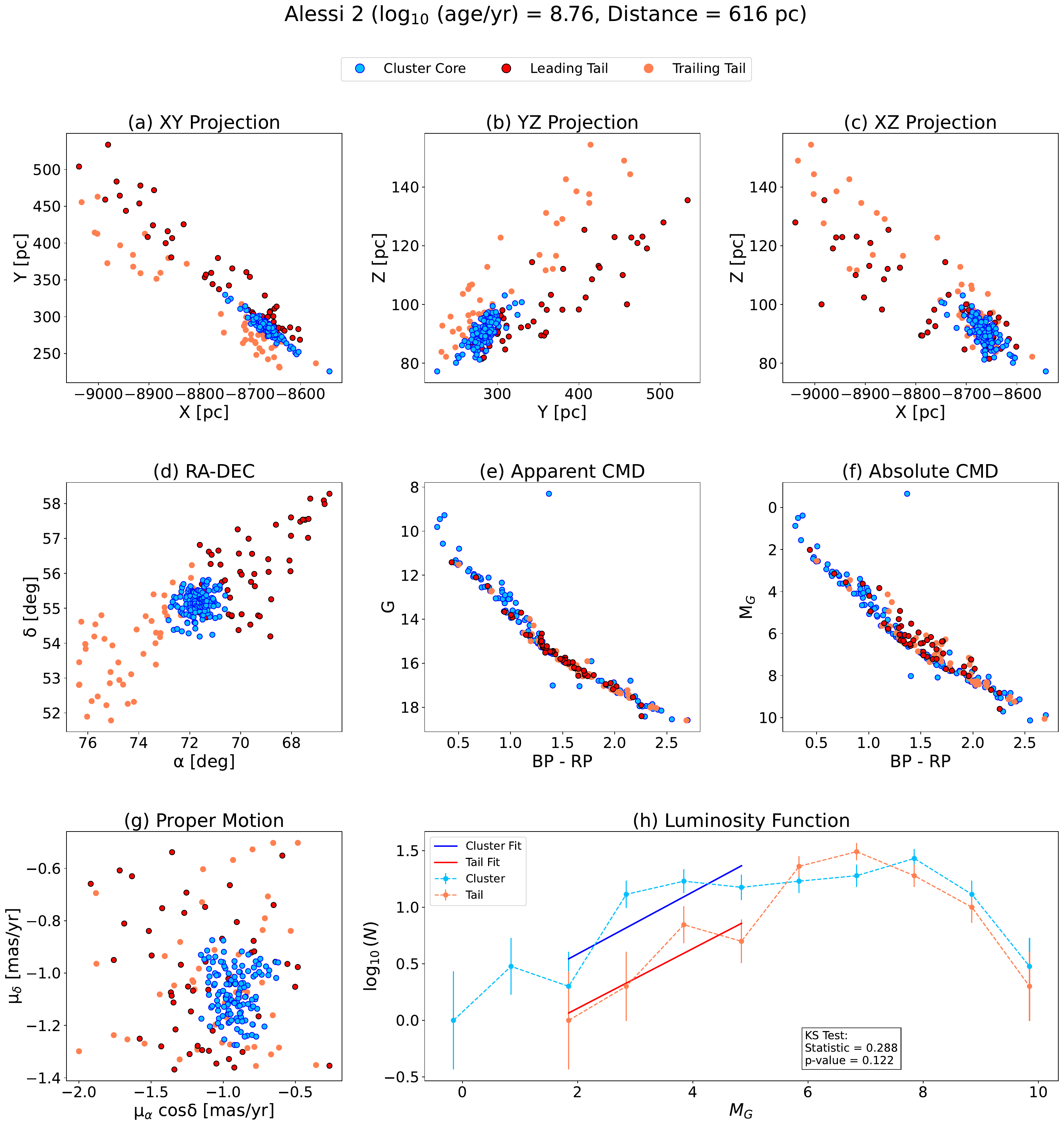} 
\caption{Same as the Fig. \ref{fig:A1} for Alessi 2.}
\label{fig:A2}
\end{figure*}

\begin{figure*}
\centering
\includegraphics[width=1\textwidth]{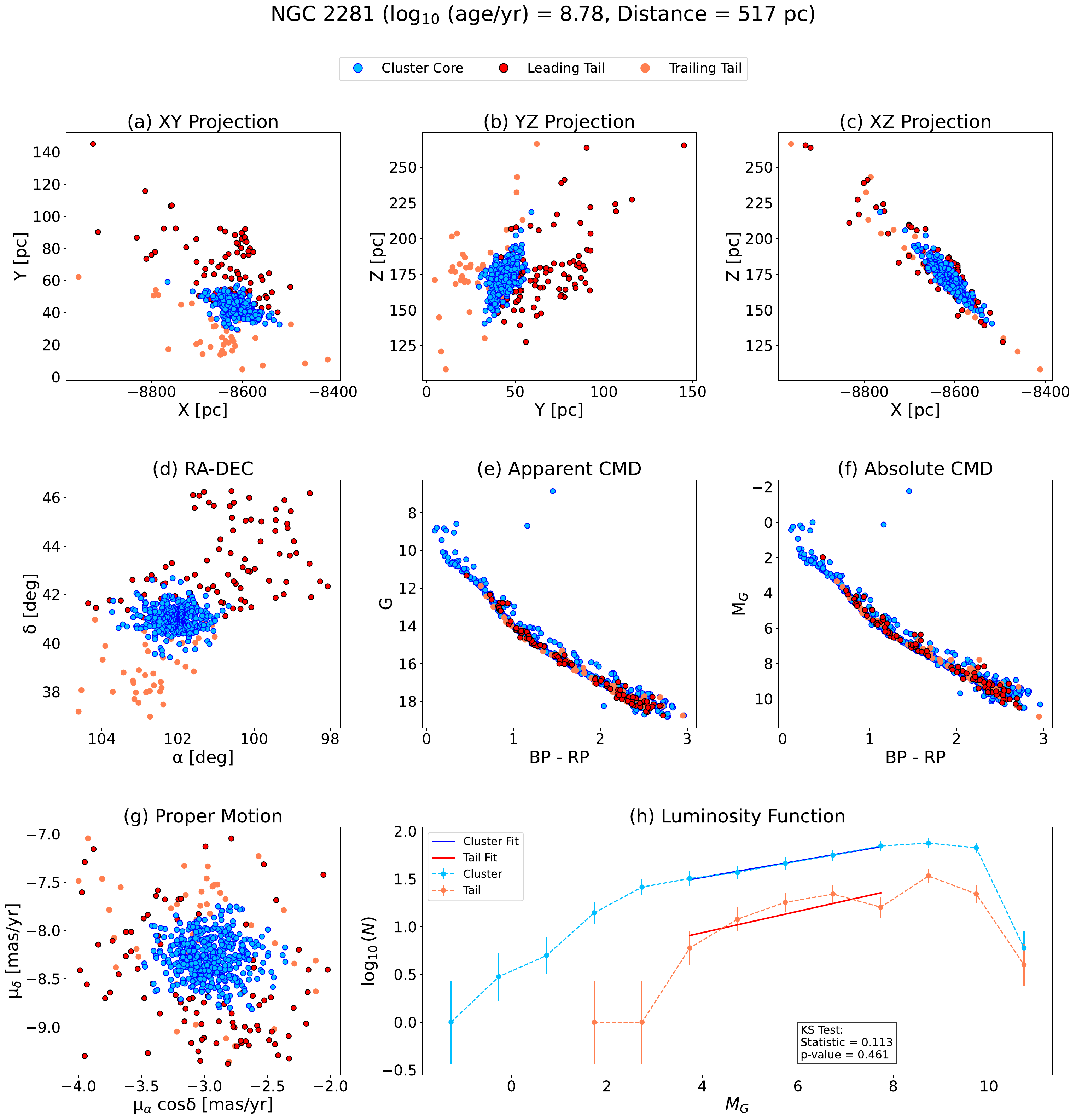} 
\caption{Same as the Fig. \ref{fig:A1}  for NGC 2281.}
\label{fig:A3}
\end{figure*}

\begin{figure*}
\centering
\includegraphics[width=1\textwidth]{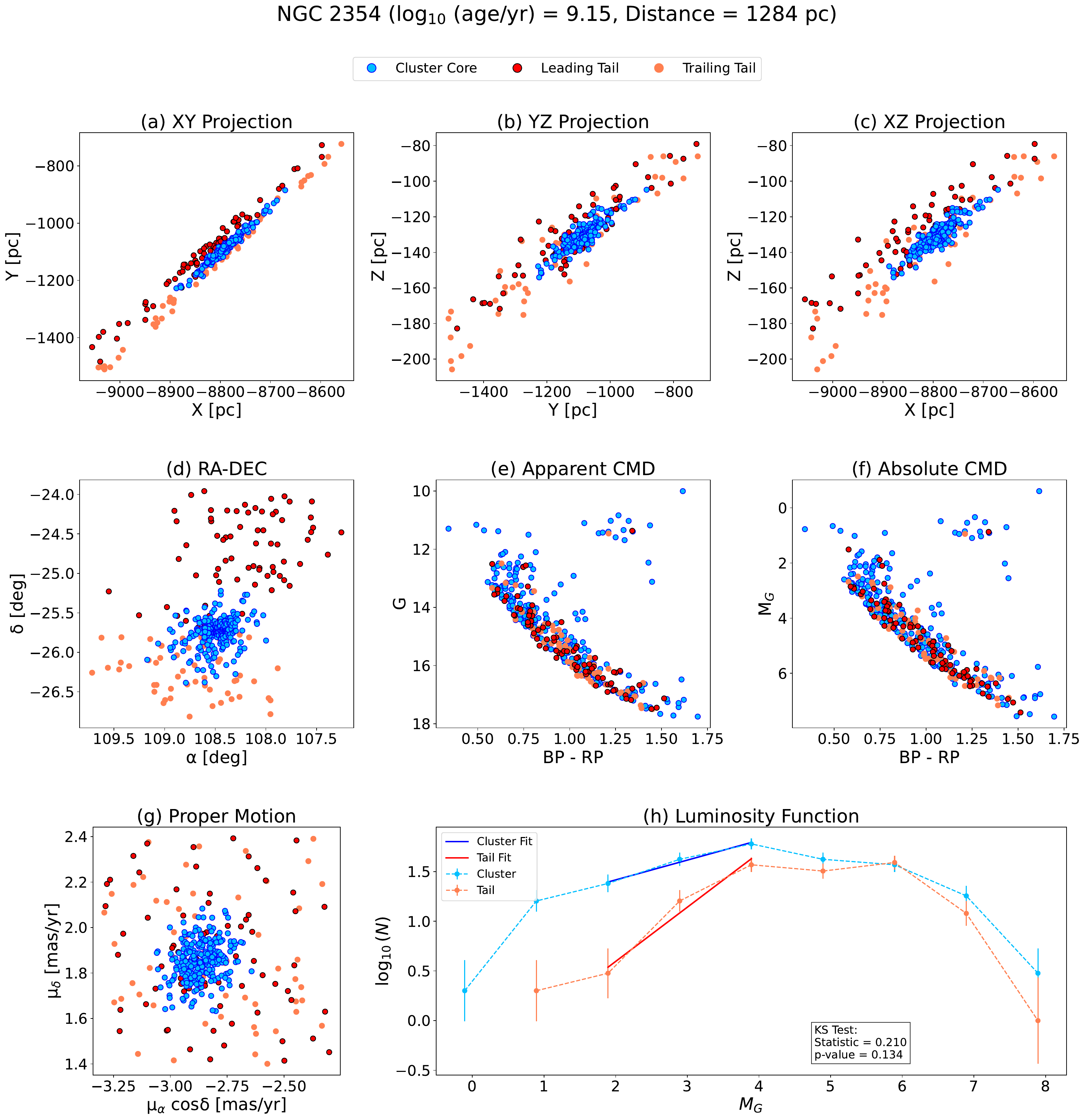} 
\caption{Same as the Fig. \ref{fig:A1}  for NGC 2354.}
\label{fig:A4}
\end{figure*}

\begin{figure*}
\centering
\includegraphics[width=1\textwidth]{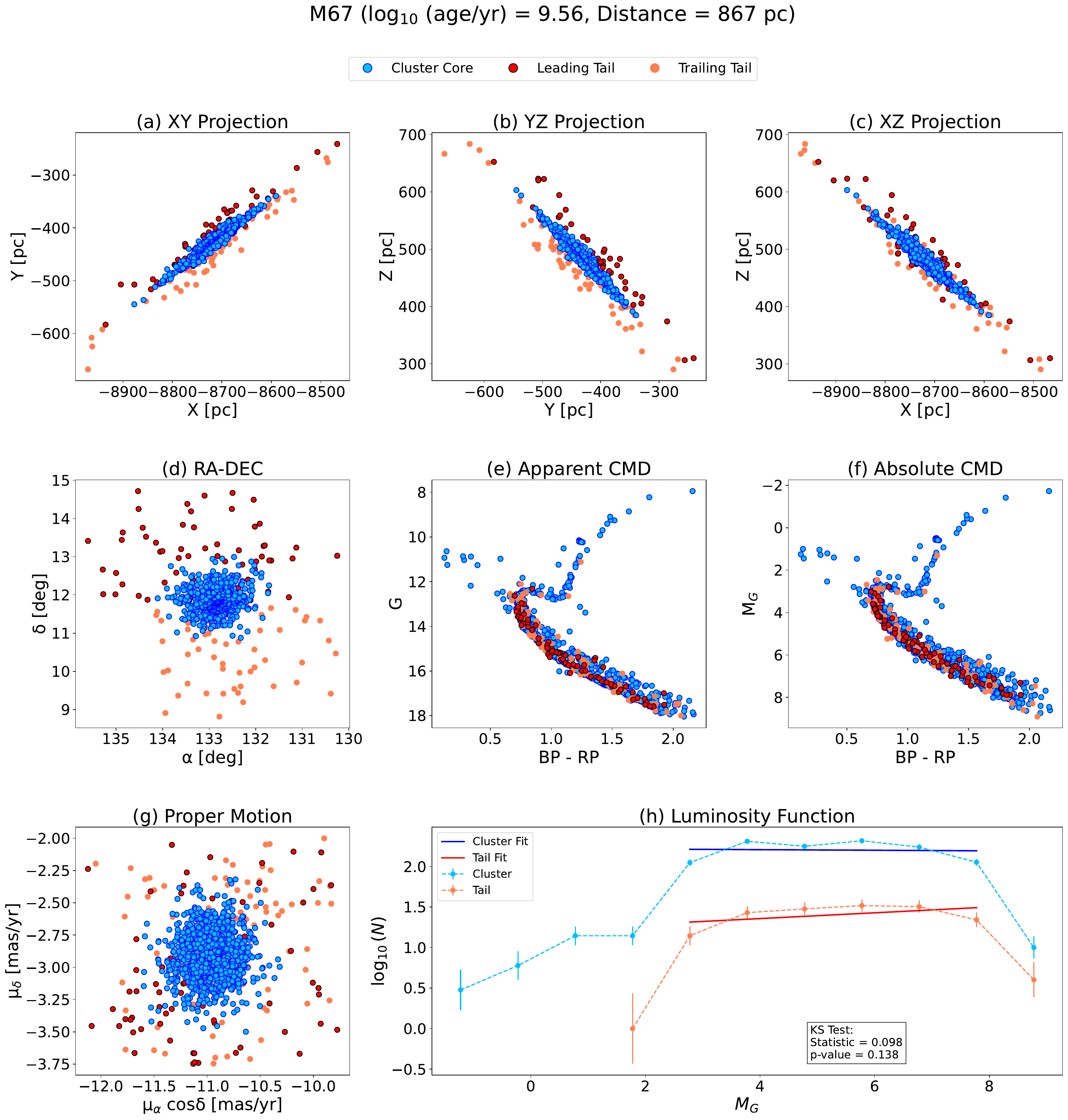} 
\caption{Same as the Fig. \ref{fig:A1}  for M67.}
\label{fig:A5}
\end{figure*}

The caption to the Appendix Figs. below \ref{fig:A1} to \ref{fig:A5} is as
follows:
(a) Spatial distribution in X-Y coordinates (b)
Spatial distribution in Y-Z coordinates. (c) Spatial distribution in X-Z coordinates. (d) Spatial distribution in equatorial
coordinates. (e) Apparent CMD. (f) Absolute CMD. (g) Vector point diagram. (h) LF (see Subsect.  \ref{subsec:luminosity}) for the cluster members (blue dots) and tails (coral dots).

\newpage
\pagebreak
\pagebreak
\pagebreak

\pagebreak

\pagebreak
\begin{figure*}
\centering
\includegraphics[width=1\textwidth]{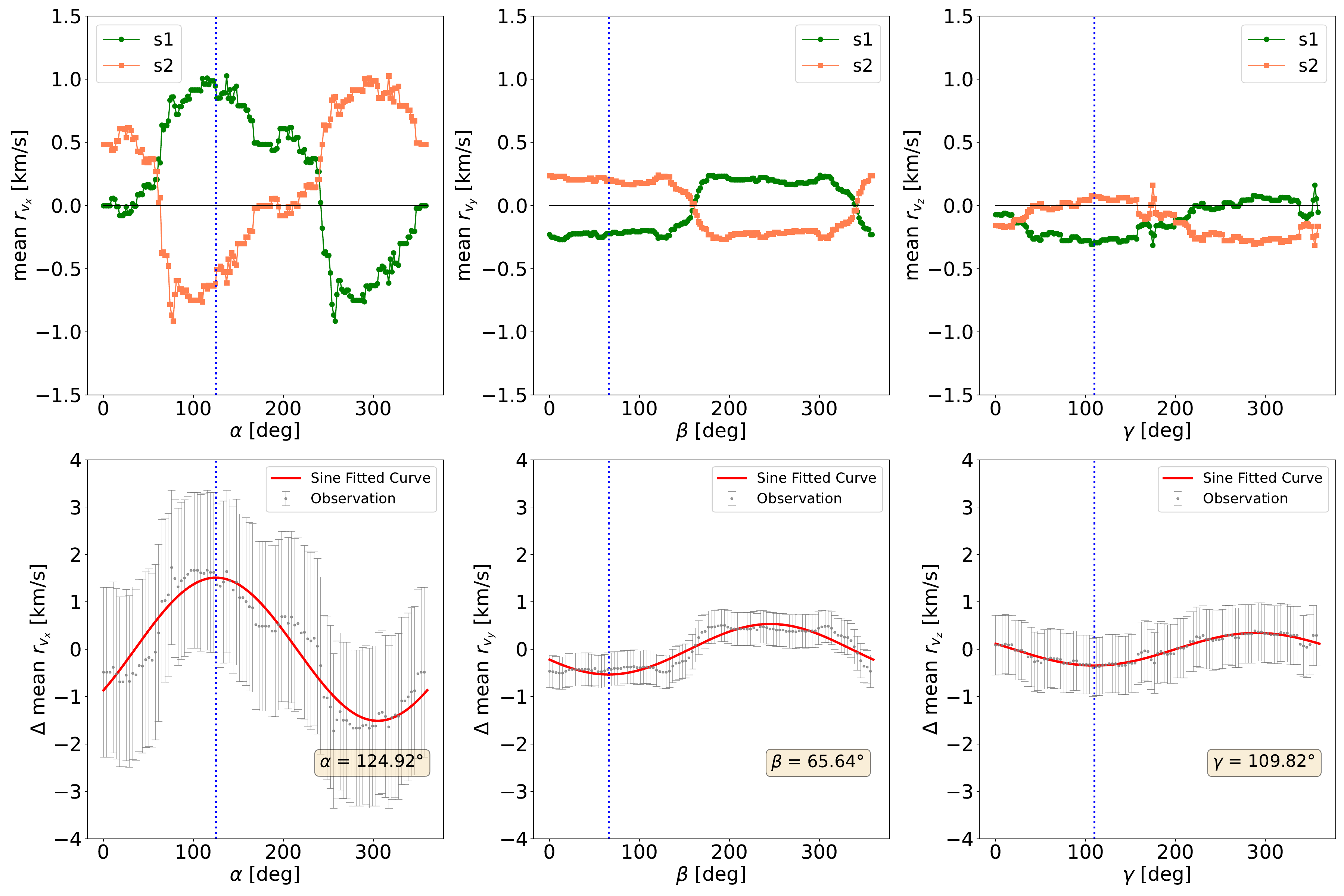} 
\includegraphics[width=0.49\textwidth]{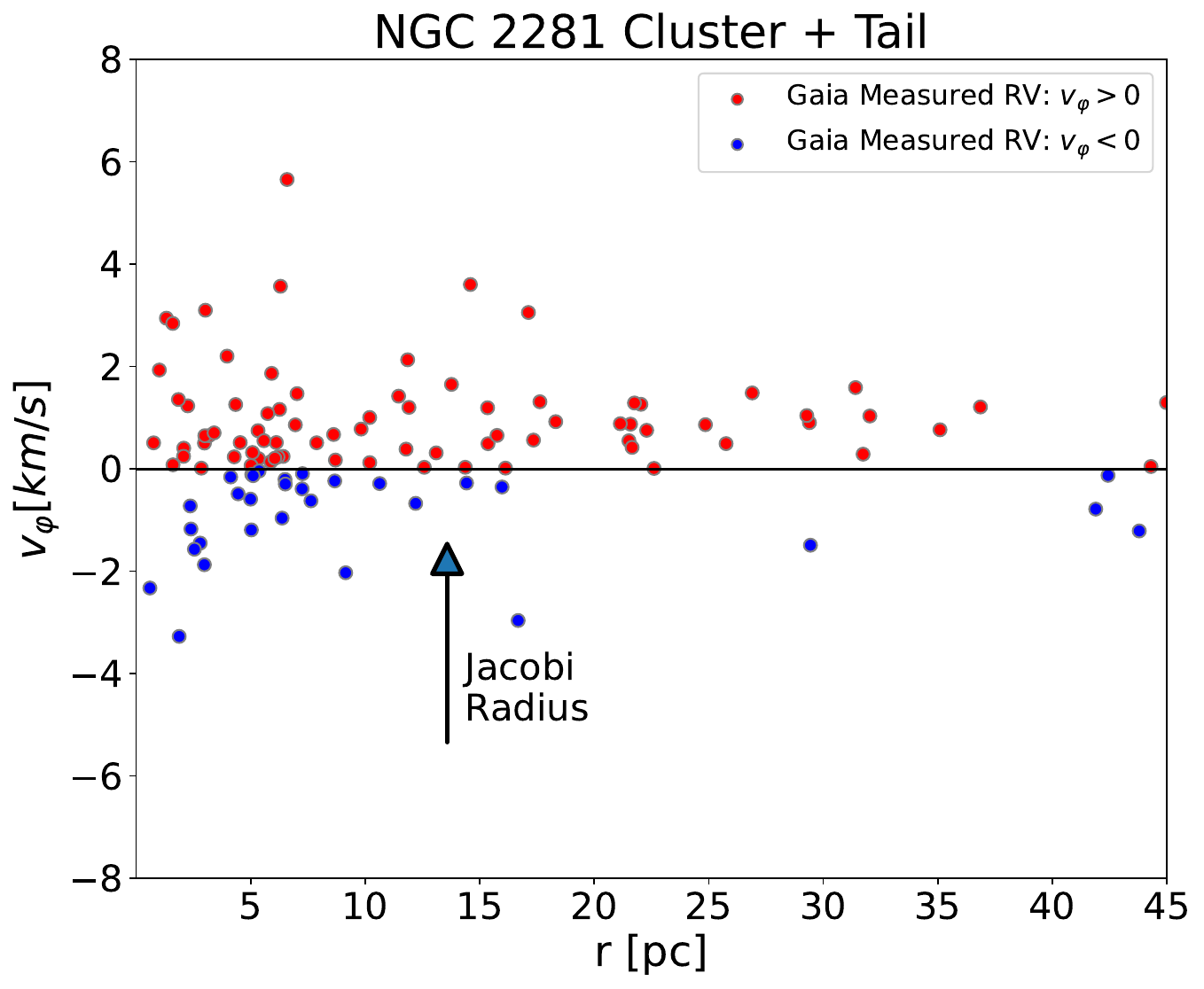} 
\includegraphics[width=0.47\textwidth]{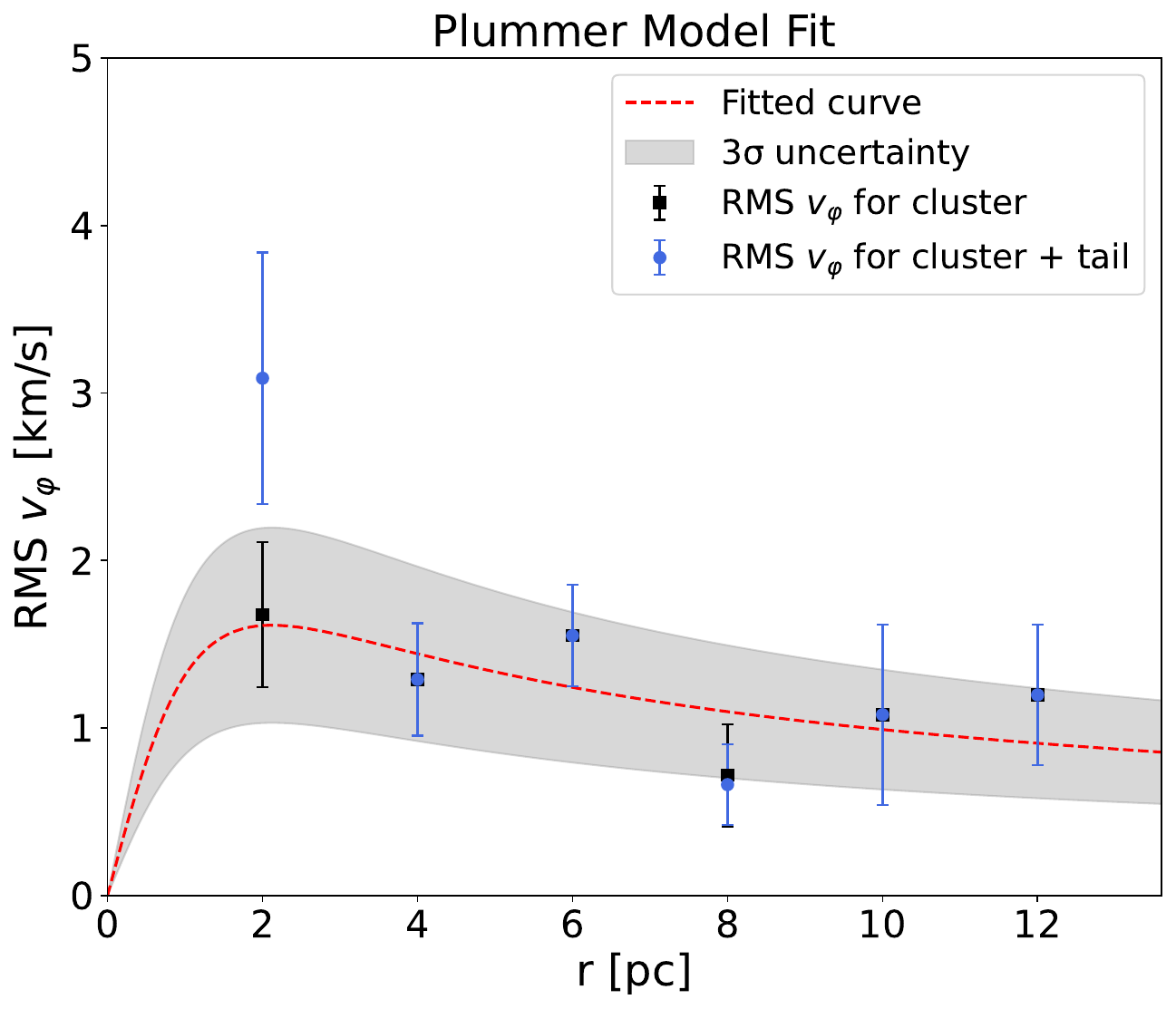} 
\caption{NGC 2281 rotation detailed in Subsect. \ref{subsec:6.4}. Red dotted line is the best-fitting Plummer model
profile, obtained using a core radius of 1.5 pc, fitted to the observed RMS rotational velocities.}
\label{fig:A6}
\end{figure*}
\pagebreak
\end{appendix}

\end{document}